\documentclass[]{aastex63}
\usepackage{graphicx}	
\usepackage{amsmath}	
\usepackage{amssymb}	
\usepackage{breqn}
\usepackage{braket}
\usepackage{chemfig}
\usepackage{float}
\usepackage{wasysym}
\usepackage{mathtools}
\usepackage{xcolor}
\usepackage{multirow}

\submitjournal{ApJ}

\shorttitle{Modelling Biosignatures: Diatomics}
\shortauthors{Thomas M.\ Cross, David M.\ Benoit, Marco Pignatari and Brad K.\ Gibson}
\graphicspath{{./}{figures/}}

\begin{document}

\title{A Large-scale Approach to Modelling Molecular Biosignatures:\\
The Diatomics}

\correspondingauthor{Thomas M. Cross}
\email{T.Cross-2019@hull.ac.uk}

\author{Thomas M.\ Cross}
\affiliation{E.~A.~Milne Centre for Astrophysics, Department of Physics and
                Mathematics, University of Hull, HU6 7RX, United Kingdom}
\author{David M.\ Benoit}
\affiliation{E.~A.~Milne Centre for Astrophysics, Department of Physics and
                Mathematics, University of Hull, HU6 7RX, United Kingdom}

\author{Marco Pignatari}
\affiliation{E.~A.~Milne Centre for Astrophysics, Department of Physics and
                Mathematics, University of Hull, HU6 7RX, United Kingdom}
\affiliation{Konkoly Observatory, Research Centre for Astronomy and Earth Sciences, Hungarian Academy of Sciences, \\ Konkoly Thege Miklos ut 15-17, H-1121 Budapest, Hungary}
\affiliation{NuGrid Collaboration, \url{http://nugridstars.org}}
\affiliation{Joint Institute for Nuclear Astrophysics - Center for the Evolution of the Elements}

\author{Brad K.\ Gibson}
\affiliation{E.~A.~Milne Centre for Astrophysics, Department of Physics and
                Mathematics, University of Hull, HU6 7RX, United Kingdom}
\affiliation{Joint Institute for Nuclear Astrophysics - Center for the Evolution of the Elements}



\begin{abstract}
This work presents the first steps to modelling synthetic rovibrational spectra for all molecules of astrophysical interest using a new approach implemented in the Prometheus code. The goal is to create a new comprehensive source of first-principles molecular spectra, thus bridging the gap for missing data to help drive future high-resolution studies. Our primary application domain is for molecules identified as signatures of life in planetary atmospheres (biosignatures), but our approach is general and can be applied to other systems.
In this work we evaluate the accuracy of our method by studying four diatomic molecules H$_2$, O$_2$, N$_2$ and CO, all of which have well-known spectra.
Prometheus uses the Transition-Optimised Shifted Hermite (TOSH) theory to account for anharmonicity for the fundamental $\nu=0 \rightarrow \nu=1$ band, along with thermal profile modeling for the rotational transitions. To this end, we expand TOSH theory to enable the modeling of rotational constants. We show that our simple model achieves results that are a better approximation of the real spectra than those produced through a harmonic approach. 
We compare our results with high-resolution HITRAN and ExoMol spectral data. We find that modelling accuracy tends to diminish for rovibrational transition away from the band origin, thus highlighting the need for the theory to be further adapted.

\end{abstract}



\section{Introduction}
\label{section:Introduction}

In the past few decades, the search for the origin of life in the Universe and the detection of its chemical signatures and primary building blocks has driven remarkable scientific achievements in astronomy and astrobiology \citep[see e.g.,][]{2008AsBio...8..715D}. A fundamental step to answer some of those questions is the observation of biologically-relevant molecules in the Universe. However, apart from extremely well studied molecules such as water \citep{97ViTePo, 18PoKyZo}, ammonia \citep{09YuBaYa, 19CoYuTe}, methane \citep{13BrSuBe,14YuTeXX} and methanol \citep{61FaWhXX,16BeFuWo}, the basic data needed to detect most biomolecules is incomplete. Existing data was gathered through meticulous experiments and computations and built from the bottom up to cover about a hundred of molecules and their ro-vibrational spectra \citep{20TeYuAl}. The molecular data has been carefully curated into well-known databases in the field such as HITRAN \citep{17GoRoHi}, ExoMol \citep {20TeYuAl}, CDMS \citep{2016JMoSp.327...95E}, JPL \citep{2005IAUS..231P.270P}, for example. Today those databases are fundamental resources for molecular line detection and form the backbone of astrochemistry and astrobiology research using data from present observation missions like ALMA \citep[][]{18HaSaMa}, HUBBLE \citep[][]{2018AJ....156..283E, 17DaMoTs}, TESS \citep{2020AJ....160..280C} and from future missions like PLATO \citep[][]{19KaNiGo}, JWST \citep[][]{15BaAiIr} and Ariel \citep[e.g.,][]{tinetti:18}, to name a few.

Molecular lines are ideal observation targets to study different astrophysical sources. Fundamental information about the properties of the early Universe can be collected from integrated galactic spectra at different redshifts \citep[e.g.,][]{muller:06, costagliola:11, aladro:15, zhang:18}, or from a number of local interstellar medium sources \citep[e.g.,][]{bacmann:12, rivilla:16, shimajiri:17}, from stars \citep[e.g.,][]{yong:03, hedrosa:13} and from both planets \citep[e.g.,][]{greaves:20, 14WeMaAt, 06TrFlFo} and exoplanets \citep[e.g.,][]{swain:09, tinetti:13, 13TeTiSa, 19GuSoBr}. 

The evolution of chemistry in the Universe has captured the increase of complexity from a metal-free environment \citep[e.g.,][]{galli:98} to a metal-, dust- and ice-rich environment \citep[e.g.,][]{wakelam:08, 08GaWeHe}. The stellar production of elements, such as oxygen silicon and iron from the first generation of core-collapse supernovae \citep[e.g.,][]{rauscher:02,nomoto:13,sukhbold:16,ritter:18} and carbon and nitrogen from low-mass stars \citep[e.g.,][]{karakas:14,cristallo:15,pignatari:16}, formed the building blocks of bio-molecules that are observed today.

Within this same context, the search for molecular fingerprints of life in atmospheres of exoplanets represents a fundamental goal of astrobiology.
A biosignature gas is defined as a gas that is produced by life and accumulates in a planet's atmosphere. An ideal biosignature would be unambiguous with living organisms being its unique source \citep[][]{20GrRiBa}. In reality, many biosignatures can be also produced through abiotic processes and can act as false positives.

For a molecule to be classified as a biosignature under the framework of \citet{16SeBaPe}, the molecule needs to fulfill a series of criteria. Here we will detail what we believe to be the three main points, further classification details can be found in the paper referenced. The first main criterion is molecular stability, identifying compounds that are stable on the order of days as a pure compound at standard conditions and if they are stable to reactions with water. The second main criterion is volatility: the likelihood a molecule would be in gaseous form at standard conditions. Volatility is difficult to assess so \citet{16SeBaPe} used boiling points instead. A molecule with a boiling below 150$^{\circ}$c was considered sufficiently volatile. The third main criterion focuses on molecular size. Only molecules of up to six non-hydrogen atoms were considered during the search. This criterion was used to limit the number of possible molecules as their number increases exponentially with each additional non-H atom. Moreover, smaller molecules are more likely to be found in gaseous form in planetary atmospheres. 

The All Small Molecules (ASM) catalogue described by \citet{16SeBaPe} contains over 14,000 biosignature molecules that comply with the criteria discussed above. For ease throughout this paper we will refer exclusively to the biosignature portion of this catalogue as the SBP, an acronym which is simply made from the first letter from each of its author's last names, as ASM also details non-biogenic molecules.

In order to be able to detect molecules in the SBP catalogue, there is a need for the laboratory astrophysics community to characterise the spectral features of each entry. Despite the simple nature of the entries in the catalogue, and considering only the rovibrational portion of the spectrum, there are thousands of biosignatures within SBP that have incorrect, incomplete or completely unknown spectra \citep[][]{19SoPeSe}.

There is therefore a pressing need for vast quantities of spectral signatures to be characterised, as shown by the very recent spectroscopic discovery of phosphine on Venus by \citet{20GrRiBa}. Not only is this molecule believed to be a biosignature but it was detected in such quantities that it cannot be currently explained abiotically \citep{20BaPeSe}. One could argue that the spectroscopic analysis which discovered phosphine would have not been possible without the efforts to produce accurate and complete computed line lists, ranging from room temperature \citep{13SoYuTe} to up to 1500K \citep{14SoAlTe}. Especially when we consider that suspected inaccuracies found in regions of previous phosphine data may have contributed to past misinterpretations of astronomical spectra \citep[][]{14SoAlTe} (as an example, the PH$_3$ database was suspected of containing some inaccuracies for the 4.5$\mu m$ line intensities, as discussed in \citet{14DeKlSa}).

Following this discovery, \citep[][]{21TrSyRo} noted the importance of being able to detect spectroscopically the presence of phosphorous bearing species. They enumerated a list of phosphorous bearing species, which could potentially be detected in planetary atmospheres, and compiled all available spectral data. As expected the data was scarce, therefore they used established computational quantum chemistry methods (CQC) to produce approximate spectra to fill in the gaps.  

Generating an entire set of SBP spectra for a given frequency range is a huge task. Indeed, experimentally it would be hard to record spectra for the entire catalogue without a concerted worldwide effort. 

Alternatively, a significant part of the SBP catalogue could be explored computationally. This is our current approach for this study, which focuses on the rovibrational spectra of diatomic molecules in SBP. Nonetheless the envisaged task requires a reliable means of producing good quality spectral data at reasonably low computational cost so that it can later be easily extended to produce spectral signatures for the rest of the catalogue. 

\begin{figure}
    \centering
    \includegraphics[width=\columnwidth]{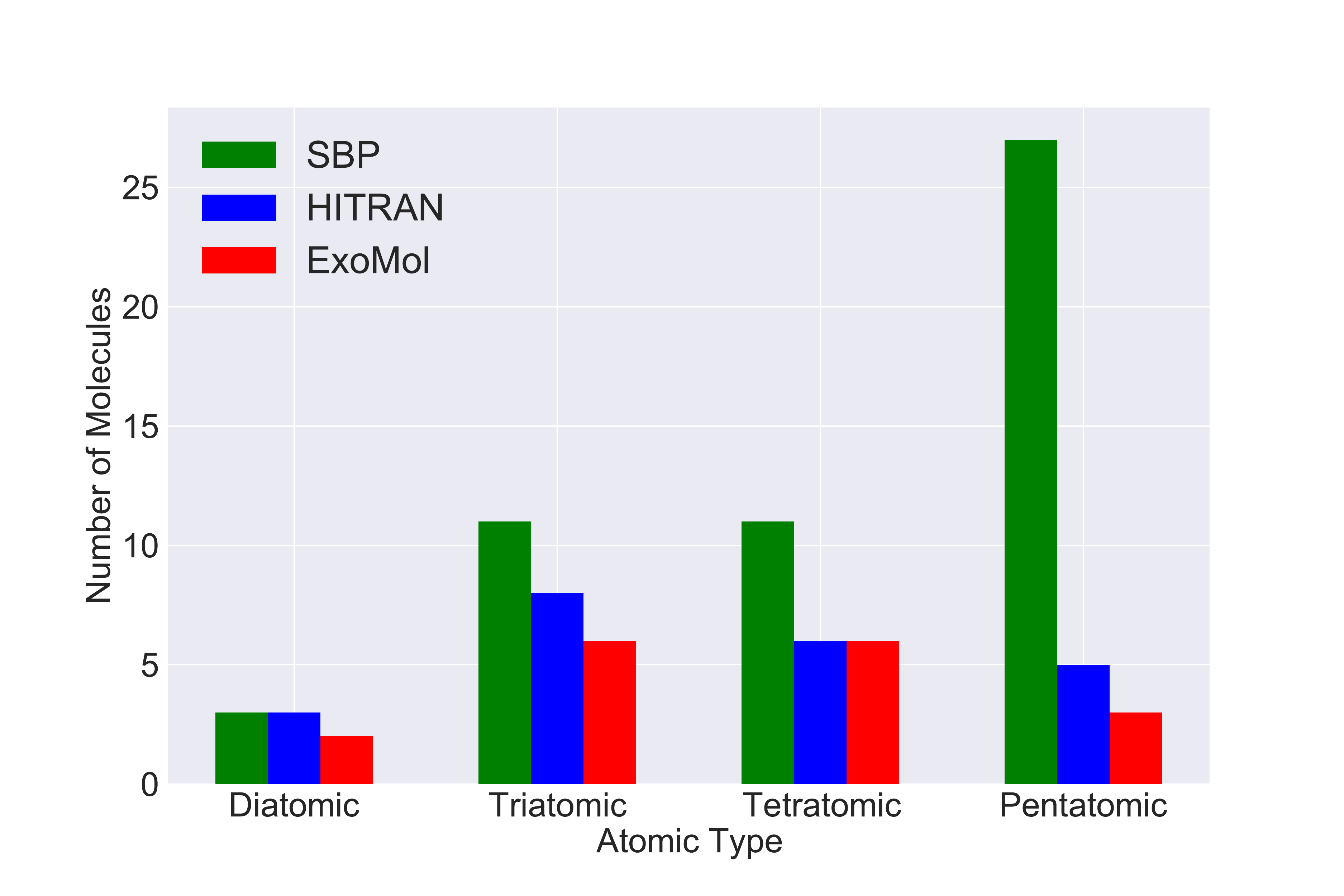}
    \caption{Comparing the availability of data from HITRAN and ExoMol with the molecules needed for the SBP. Only up to pentatomic molecules has been included here.}
    \label{fig:Comparing_catalogues_up_to_pentatomic}
\end{figure}

The current publicly-available version of HITRAN contains detailed spectra for 49 molecules and also atomic oxygen. A further description of HITRAN is provided in section \ref{subsection:comparison_catalogues}. Out of those 49 molecules, only 26 are actually believed to be biosignatures according to SBP. To break this down even further, Figure \ref{fig:Comparing_catalogues_up_to_pentatomic} compares the number of biogenic molecules for each total number of atoms, up to pentatomics. Rounding the SBP down to exactly 14,000 biosignatures, gives HITRAN a completion rate of roughly 0.19\%.

The current version of ExoMol (see also further background in section \ref{subsection:comparison_catalogues}) contains high-quality spectra at a large range of temperatures for 81 molecules, however only 18 of these molecules are biogenic. We report in Fig.~\ref{fig:Comparing_catalogues_up_to_pentatomic} a similar breakdown to the one done for HITRAN. Here we see that ExoMol has a completion rate of approximately 0.13\%.

Combining both databases gives a maximum total of 26 molecules (excluding SBP entries overlaps). Assuming a constant rate of growth for these databases, without significant change to available technology and resources, it will be difficult to produce spectra for even 1\% of SBP, let alone the entire catalogue. 

The work described here provides a new approach that is meant to complement and support the high-resolution databases mentioned above.
We aim to 
obtain spectra from first principles, but 
faster than existing approaches (such as VSCF/VCI shown by \citet{19ClBeXX} or DVR shown by \citet{19McMaHo} for example) possibly at the expense of accuracy. 

Primarily we are trying to address the gap between approximate, fundamental-only models (such as harmonic, non-adapted TOSH or RASCALL), and labour intensive, extremely accurate, line lists (such as ExoMol and HITRAN). This is done by developing a method which borrows key physics from both approaches to create an intermediate approach. We present here a model that approximates the fundamental ($\nu=0\rightarrow\nu=1$) band \emph{and} its rotational profile, computed at 300K. The intensities for the spectra use thermal population equations, and as such have the additional ability of modelling rovibrational lines at differing temperatures. The investigation into the effect of differing temperatures on spectra is not considered in this paper.

Additionally our approach is designed to be simple, open-source, and computationally cheap, but still more detailed than the aforementioned approximate fundamental-only methods. For example, we do not currently model effects such as external electric or magnetic fields (see, for example, PGOPHER \citep[][]{16WeXXXX}), to preserve simplicity and low computational cost.

Once the catalogue is constructed it should, and will, act as a living document, preferably being continually updated with better results. The rationale for this approach is to provide a wide coverage of the SBP database to help with both detection and atmospheric models with a reasonable accuracy that could be improved with further releases of the database.

In this study we have applied our work to the diatomic biosignature molecules, of which there are three in the SBP catalogue. We have also additionally included a fourth diatomic, carbon monoxide (CO), for reasons discussed in section \ref{subsection:ResultsCO}.

The paper is organised in the following manner: in section \ref{section:Methodology}, we introduce the methodology and theory required to produce the synthetic data. The following section, section \ref{section:Results}, presents and then discusses our results. We conclude our findings in section \ref{section:Conclusions}.

\section{Methodology}
\label{section:Methodology}
\subsection{Requirements}
When modelling the rovibrational spectrum, we considered two key parameters: the band origin and the rotational constants. The band origin crucially determines the position of its corresponding rotational lines and thus has a strong influence on the overall shape of the rovibrational band. In our model, we have assumed this aspect to be the most important.

The rotational constants have an effect on the spacing of the transitions on each of the rotational branches. As a secondary effect, anharmonicity causes the rotational constant to depend on the vibrational state. Indeed, the bond lengthening in excited vibrational states allows a Q-branch progression (a set of purely vibrational transitions with no rotational transitions, see eq.~\ref{eq:28}), since the rotational constants of both starting and final vibrational states are different. 
An identical rotational constant for all vibrational level (such as predicted by a simple harmonic oscillator approximation, for example) incorrectly suggests that vibrations and rotations are independent. This leads to the Q branch transitions bunching up at a single position, rather than displaying a typical progression. Therefore, for an accurate approximation for the entirety of the spectrum, the rotational constant will need to vary with vibrational level \citep[][]{01PaJoOl}.


The transition-optimised shifted Hermite (TOSH) theory is a modern approach to treating the nuclear wave function that does predict correctly most band origins \citep{08LiGiGI}. TOSH is effectively a simplification of second-order vibrational perturbation theory (VPT2), and applies limited-order perturbation theory to the nuclear Schr{\"o}dinger equation. By doing this, it avoids any degeneracy issue that may arise from standard second order vibrational perturbation theory \citep{90WiHaGr,06VaStXX}, for example.

Fundamentally TOSH works by introducing a shifting parameter ($\sigma$) which is used to shift the harmonic basis functions from their equilibrium position. This shift is optimised for the vibrational transition energy expansion, specifically for the fundamental vibrational transition, and as such does an excellent job at correcting the band origins (this is shown by our results in section \ref{section:Results} and Table \ref{table:comparing_spec_values}). An expression for the magnitude of the shift is described in equations \ref{eq:2}-\ref{eq:10} in section \ref{subsection:TOSH}. We have discovered however that this shift can be applied to the equilibrium distance to recover the anharmonic effects of the first vibrationally excited level (see \ref{subsection:rot_constants}).

Despite its approximate nature, TOSH is roughly a factor of two cheaper than VPT2, with only some loss in accuracy as a compromise \citep[][]{08LiGiGI}. This approach also has the potential to be of reasonable accuracy for a large ensemble of molecule \citep[][]{19HaXXXX} and is straightforward to extend to larger systems \citep[][]{08LiGiGI}. 

\subsection{TOSH Theory}
\label{subsection:TOSH}
We will now briefly discuss the necessary theory required to understand the process to produce the Prometheus spectra.

For a diatomic molecule, the Hamiltonian with up to fourth order terms (quartic potential) and using mass-weighted displacement coordinate, $Q=\sqrt{\mu}x$, can be expressed as:

\begin{equation} \label{eq:1}
    \hat{H} = -\frac{1}{2}\frac{\partial^2}{\partial Q^2} + \frac{1}{2!}\eta_{ii}Q^2 + \frac{1}{3!}\eta_{iii}Q^3 + \frac{1}{4!}\eta_{iiii}Q^4
\end{equation}{}

Now imagine a shift, $\sigma$ along the coordinate $Q$. If the centre of the wave function is shifted by $\sigma$, the shape will remain the same but the anharmonic correction can be incorporated into the wave function. 

The shifted wave functions for TOSH, are now different to the harmonic wave functions, and are described by:
\begin{equation} \label{eq:2}
    \psi_n = \left(\frac{\omega^{1/2}}{\pi^{1/2}2^n n!}\right)^{1/2} e^{-\frac{\omega}{2}(Q - \sigma)^2}H_n\left[(Q-\sigma)\omega^{1/2}\right]
\end{equation}

where $H_n(x)$ naturally refers to the Hermite polynomial, and $\omega$ describes the harmonic frequency (defined using TOSH constants in equation \ref{eq:13}, see below).

The energy of this ground vibrational state is:
\begin{equation} \label{eq:3}
    E_0 = \braket{\psi_0| \hat{H} | \psi_0}
\end{equation}

\begin{dmath} \label{eq:4}
    E_0 = \frac{1}{4}\omega
    +\frac{1}{2!}\eta_{ii} \left[\frac{1}{2\omega} + \sigma^2\right]
    +\frac{1}{3!}\eta_{iii}\left[\frac{3\sigma}{2\omega} + \sigma^3\right]
    +\frac{1}{4!}\eta_{iiii} \left[\frac{3}{4\omega^2} + \frac{3\sigma^2}{\omega} + \sigma^4\right]
\end{dmath}{}

The equation for the energy of the first excited vibrational level is:
\begin{equation} \label{eq:5}
    E_1 = \braket{\psi_1 | \hat{H} | \psi_1}
\end{equation}
\begin{dmath} \label{eq:6}
    E_1 = \frac{3}{4}\omega + \frac{1}{2!}\left[\frac{3}{2\omega} + \sigma^2\right]\eta_{ii} + \frac{1}{3!}\eta_{iii}\left[\frac{9\sigma}{2\omega} + \sigma^3\right] + \frac{1}{4!}\eta_{iiii}\left[\frac{15}{4\omega^2} + \frac{9\sigma^2}{\omega} + \sigma^4\right]
\end{dmath}

Therefore the energy difference between the first vibrational state and the ground state is:
\begin{align}\label{eq:7}
    \Delta E^{TOSH} &= \braket{\psi_1^{\ast} | \hat{H} | \psi_1} - \braket{\psi_0^{\ast} | \hat{H} | \psi_0}
    \\
    \label{eq:8}
    &= \omega + \frac{\eta_{iiii}}{8\omega^2} + \frac{\eta_{iii}\sigma}{2\omega} + \frac{\eta_{iiii}\sigma^2}{4\omega}
\end{align}

Which can now be compared with the energy from unshifted wave function obtained through second-order perturbation theory, VPT2.

\begin{equation}\label{eq:9}
    \Delta E^{VPT2} = \omega + \frac{\eta_{iiii}}{8\omega^2} - \frac{5\eta_{iii}^2}{24\omega^4} - \frac{\eta_{iiii}^2}{32\omega^4}
\end{equation}

Within the TOSH theory derivations, $\sigma$ is assumed to be small and therefore the $\sigma^2$ term can be neglected. By comparing the coefficient of $\eta_{iii}$ in both TOSH (\ref{eq:8}) and VPT2 (\ref{eq:9}) expressions, a suitable value for the shift can be obtained:
\begin{equation}\label{eq:10}
    \sigma = -\frac{5}{12} \frac{\eta_{iii}}{\omega^3}
\end{equation}

Due to its derivation, this value of the shift parameter is only optimal for the $0 \rightarrow 1$ transition. It is also worth noting that the original TOSH paper of \citet{08LiGiGI} does not match the $\eta_{iiii}$ terms for the VPT2 energy expression, but remedied this by stating this term is often neglected anyways. Many intermediate processes for the derivations above have been assumed and therefore omitted. For a full derivation, please see the appendix, section \ref{appendix:energy_deriv}. 

One lesser-considered aspect of having a displaced vibrational wave function (mimicking what happens for the exact wave function) is the usage the shift parameter in order to include anharmonic effects in the computation of rotational constants. We explore this possibility in section \ref{subsection:rot_constants}.

\subsection{Potential Energy Curve, or PEC}
\label{subsection:PEC}
As shown in equations \ref{eq:8} and \ref{eq:10}, the TOSH approach (and our Prometheus implementation, see (\ref{subsection:Prometheus})) only requires a selection of anharmonic constants for the chosen diatomic. These constants can be obtained either through finite-difference methods or by computing the derivatives of the PEC directly, if a functional form is available. In this study, we do not focus on producing PECs, but we use existing ones to derive the required anharmonic constants using a quartic fitting procedure described below.

Prometheus uses the numerical points that make up the PEC to perform a quartic fit (see eq.\ref{eq:11}) and extract the anharmonic constants. We use this methodology for two reasons. Historically quartic force fields have been used as a generic polynomial form to model PECs. In particular it is used as a tool for analyzing and producing rovibrational spectra for molecules of interest to astrophysical observation \citep[][]{19FoLeXX}. Secondly, the TOSH theory requires only the 2nd, 3rd and 4th order derivatives, without any need for higher-order derivatives.

The quality of the PEC is paramount and will affect the accuracy of the TOSH constants derived from our quartic fit. The code performs a local fit, centered around the equilibrium bond length, rather than a global fit which would include the potential wall. 
The local fit is determined via an ``inclusion range" determined by starting with the equilibrium bond length and including points at differing percentages of the equilibrium bond length. The ranges tested within the code span from $\pm20\%$ to $\pm7\%$ of the equilibrium bond length, in intervals of $1\%$. 

The fit accuracy has been tested for each range using a simple chi-square test. The range with the lowest chi-square value, or the first range with a sufficiently low chi-square value (arbitrarily defined as $1\times10^{-7}$) was selected. This was to ensure the range providing the best fit (while still considering the point density of the remaining curve) to the reference data is selected and used to ultimately produce a rovibrational spectrum. This method has been chosen to ensure a consistent selection of inclusion ranges for the potential energy curve of each molecule. It also removes any bias we could have introduced into the analysis by selecting the ranges ourselves.

The quartic fit applied to the PEC is described by equation:
\begin{equation}\label{eq:11}
    V = E_0 + \frac{1}{2!}\zeta_{ii}X^2 + \frac{1}{3!}\zeta_{iii}X^3 + \frac{1}{4!}\zeta_{iiii}X^4
\end{equation}
Where $X=r-r_e$ is a displacement coordinate, with $r_e$ describing the equilibrium bond length. Note: $E_0$ has been included to help Prometheus with the fitting by centering around 0, and has little importance beyond this.

The fit only returns $\zeta$ constants, $\zeta_{ii}$, $\zeta_{iii}$ and $\zeta_{iiii}$. Which need to be massed weighted for them to become the TOSH constants, as described in the Hamiltonian in equation \ref{eq:1}. This was done in the following manner:
\begin{equation}\label{eq:12}
    \eta_{ii} = \frac{\zeta_{ii}}{\mu},\quad  \eta_{iii} = \frac{\zeta_{iii}}{\mu^{3/2}},\quad \eta_{iiii} = \frac{\zeta_{iiii}}{\mu^2}
\end{equation}

Note that fits of order higher than quartic lead to sizable errors and inaccurate results. Additionally, we used spline interpolation to ensure the data spans the entire range specified.

\subsection{Choice of a PEC for diatomic anharmonicity constants determination}

In this study, we obtain the anharmonicity constants by computing the derivatives of a quartic fit to a given PEC. As previously alluded to, these constants are the only data required for Prometheus to operate. The technique used to derive these constants (finite-difference, derivatives of PEC etc.) are inconsequential, but require a suitable description of the inter-atomic potential around the equilibrium bond length in order to produce reliable spectra. In order to avoid issues originating from a PEC of sub-spectroscopic quality, we choose to use only data that has been validated through spectroscopic comparisons. 

Indeed, while a number of PEC for diatomics are available in the literature, some of the molecules in our selected set have a particularly challenging electronic structure (CO, N$_2$ and O$_2$). This in turn affects the quality of the anharmonicity constants, which then impacts accuracy the ro-vibrational lines. As our study focuses on the quality of the ro-vibrational approach, rather than the quality of the electronic structure approach used for the potential, we chose well-reported modern PECs, obtained through either RKR inversion (see \citet{47ReXXXX} for further details) or ab initio calculations. All the potentials used within this study are for the ground state of their respective molecule.

The potential energy curves used are those from \citet{87ScLeXX} (the Mass-Independant Clamped Nuclei, $V_{BO}$, potential with adiabatic correction, $\Delta V_{ad}$)  for the $^1\Sigma_g^+$ state of H$_2$. The potential energy curve for $^1\Sigma^{+}$ for CO is from \citet{18MeStEr} (specifically the CO\_X\_Func\_edited Born-Oppenheimer, UBO, potential located within the supplementary material). The potential energy curves for the $^3\Sigma^{-}_{g}$ state of O$_2$ from \citet{14YuDrMi} and the potential for the $^1\Sigma^{+}_{g}$ state of N$_2$ from \citet{93EdRoLa}.

\begin{table}[]
\begin{center}
\begin{tabular}{c c c c c }
\hline
\hline
Constant & H$_2$       & O$_2$       & N$_2$       & CO         \\ \hline
{$\eta_{ii}$ ($E_h a_0^{-2} m_e^{-1}$)}   & $4.040\times 10^{-4}$  & $5.183\times 10^{-5}$  & $1.156\times 10^{-4}$  & $9.803\times 10^{-5}$  \\ 
{$\eta_{iii}$ ($E_h a_0^{-3} m_e^{-3/2}$)}  & $-4.971\times 10^{-5}$ & $-1.718\times 10^{-6}$ & $-4.048\times 10^{-6}$ & $-3.430\times 10^{-6}$ \\ 
{$\eta_{iiii}$ ($E_h a_0^{-4} m_e^{-2}$)} & $5.317\times 10^{-6}$  & $4.789\times 10^{-8}$  & $1.069\times 10^{-7}$  & $8.928\times 10^{-8}$  \\ 
\hline\hline
\end{tabular}
\caption{Anharmonic constants computed by Prometheus (TOSH) for each molecule. Atomic units are used throughout, where $E_h$, $a_0$ and $m_e$ represent Hartrees, Bohr radius and electron rest mass respectively. The constants are derived from a quartic fit to the numerical potential energy curves (see sec.\ref{subsection:PEC}) for details.}
\label{table:comparing_eta}
\end{center}
\end{table}

H$_2$ is an ab initio potential, CO a semi-empirical potential, while N$_2$ and O$_2$ are spectroscopic RKR curves. This has been done in part to highlight the flexibility of our model, which can accept any type of PEC, rather than being bound to a single format. Naturally, if the quality of the curve is low, the ro-vibrational results will reflect this fact, but will still remain a better approximation than a pure harmonic approach.

The necessary anharmonicity or TOSH (quartic) constants, obtained using optimal quartic fitting ranges, for each molecule in this study are given in table \ref{table:comparing_eta}.

\subsection{Spectroscopic Constants}
\label{subsection:Spectroscopic_Constants}
The TOSH framework allows us to determine key constants, such as the harmonic frequency, $\omega$, the anharmonic fundamental transition, $\Delta E^{TOSH}$ or $\nu_{0\rightarrow 1}$, and the coordinate shift, $\sigma$.

The value of $\omega$ is calculated using:
\begin{equation}\label{eq:13}
    \omega = \left(\eta_{ii}\right)^{1/2}
\end{equation}

The value of $\Delta E^{TOSH}$ is obtained from eq.~\ref{eq:8}, and the shift, $\sigma$, is given by eq.~\ref{eq:10}. The associated errors for each spectroscopic constant is also calculated to allow an assessment of the error on the positions of transitions in the spectra produced by Prometheus (see \ref{subsection:Prometheus}).

\subsection{Rotational Constants}
\label{subsection:rot_constants}
The state-specific rotational constant, $B_v$, for a diatomic molecule is defined as:
\begin{align}\label{eq:14}
    B_v &= \frac{h^2}{8\pi^2c\mu r_v^2}
\end{align}
Where $v$ refers to the vibrational level, $\mu$ is the reduced mass and $r_v$ is a bond length that depends on the vibrational energy level considered. We can see that the bond constant has a reciprocal squared relationship to the bond length, meaning as length increases the constant will decrease. This is one of the fundamental equations used throughout Prometheus to ultimately create the rovibrational spectra.

In the harmonic approximation, $r_v$ is independent of the vibrational energy level and thus $r_v=r_e$, the equilibrium bond length. Consequently, in that approximation $B_v = B_e$, a fixed rotational constant obtained from $r_e$. More generally, the value of the bond length for the rotational constant for each vibrational state can be obtained from the expectation value of the position, $r_v=\langle \psi_n|r|\psi_n\rangle$. Using this expression for the harmonic model, leads to the same conclusions as earlier: $r_v=r_e$, since the harmonic wave function is symmetric around $r_e$. 

If we now consider the TOSH model, the shifted TOSH position, $x$, can be described using the harmonic position, $u$, and the shift constant, $\sigma$:
\begin{equation}\label{eq:15}
    x = u + \sigma
\end{equation}
As shown earlier, we can take the expectation value of the position, but this time implementing TOSH:
\begin{equation}\label{eq:16}
    \braket{\psi_{n}^{T}|x|\psi_{n}^{T}} = \braket{\psi_{n}^{H}(u)|u+\sigma|\psi_{n}^{H}(u)}
\end{equation}
Where $\psi_{n}^{T}$ and $\psi_{n}^{H}$, represent the TOSH (equation \ref{eq:2}) and harmonic wave functions, respectively.
\begin{equation}\label{eq:17}
   \braket{\psi_{n}^{T}|x|\psi_{n}^{T}} = \braket{\psi_{n}^{H}(u)|u|\psi_{n}^{H}(u)} + \braket{\psi_{n}^{H}(u)|\sigma|\psi_{n}^{H}(u)}
\end{equation}
The first expectation value is the position expectation position for a harmonic wave function (i.e.\ zero) and thus we can see that the expectation of the position for TOSH is simply:
\begin{equation}\label{eq:18}
    \braket{\psi_{n}^{T}|x|\psi_{n}^{T}} = \sigma
\end{equation}
The full derivation, for both required levels, can be found in the appendix section \ref{appendix:posexpval}.

Thus the TOSH model leads to $r_v = r_e + \sigma$. As is the case for the harmonic model, the TOSH approach only produces a single static value for all rotational constants (see section \ref{appendix:posexpval}). 

We cannot accurately reproduce rotational constants that exhibit rotation-vibration interaction, using solely either a TOSH or a harmonic model: a hybrid theory is required. The approach we use in Prometheus, approximates the ground-state rotational constant using the harmonic value. The first excited vibrational level uses the modified TOSH bond length derived above, where sigma is combined with the equilibrium bond length. The vibrationally-averaged bond distances are thus expressed as follows:
\begin{align}\label{eq:19}
    r_0 &= r_e 
    \\
    \label{eq:20}
    r_1 &= r_e + \sigma
\end{align} 
where $\sigma$ is not mass weighted to harmonise units.

The harmonic approach effectively costs nothing (using $r_0 = r_e$), so can be exploited to provide the ground state rotational constant with little effort.
Our approach does not necessarily provide a strong quantitative agreement but is qualitatively correct - bond length gets larger with increased vibration, hence the rotational constant changes (gets smaller). 

\subsection{Rovibrational Spectra}
\label{subsection:Rovibrational_Spectra}
In order to generate a ro-vibrational linelist and corresponding spectrum, we compute the maximum allowed quantum rotational number $J$ for a given input temperature. 
In the present study, the temperature was set to 300K to match the other data sets. The spectra from each of the experimental databases are all set at 300K also.

Rotation symmetry also influences the spectrum the molecule will produce and therefore 
conditional arguments within the intensity calculations have been created to account for this. If a diatomic molecule has a center of symmetry, hence meaning it is homonuclear, it is Raman active. This occurs instead of IR activity due to having no dipole moment, instead stretching and contraction of the bond leads to changes in the polarizability. This type of molecular response implies different transition selection rules leading to the O, Q, S branches, rather than the typical P and R branches in infrared. 

Nuclear spin also has an effect on the spectra produced. This will be discussed further in section \ref{section:Results}. The code is automated to account for additional effects by using the initial inputs of the mass of a molecule.

We use a Boltzmann distribution to obtain the relative intensities for each transition. This statistical law states that, for a system of N total molecules, only a fraction $\frac{N_J}{N}$ will occupy particular energy level $E_J$ a fraction. This can be written as:
\begin{equation}\label{eq:21}
    \frac{N_J}{N} = \frac{g_J e^{\left(\frac{-E_J}{kT}\right)}}{f}
\end{equation}
Where, the degeneracy, $g_J$, the energy of a given level, $E_J$, and the partition function, $f$ are described as:
\begin{equation}\label{eq:22}
    g_J = (2J+1)
\end{equation}
\begin{equation}\label{eq:23}
    E_J = B_0J(J+1)
\end{equation}
\begin{equation}\label{eq:24}
    f = \sum g_J e^{\left(\frac{-E_J}{kT}\right)}
\end{equation}

Hence, the full equation now becomes:
\begin{equation}\label{eq:25}
    \frac{N_J}{N} =\frac{(2J+1) e^{\left(\frac{-BJ(J+1)}{kT}\right)}}{f}
\end{equation}

From here the code calculates the transition positions using the appropriate equations depending on the rotational branches considered. The full derivations for these equations have been included in the appendix, section \ref{appendix:branches}. They are summarised as follows:
\begin{align}
    \label{eq:26}
    \nu_O &= \nu_{0\rightarrow 1} + B_1J'(J'+1) - B_0(J'+2)(J'+3)
    \\[10pt]
    \label{eq:27}
    \nu_P &= \nu_{0\rightarrow 1}
    - (B_1 + B_0)(J'+1) + (B_1 - B_0)(J'+1)^2
    \\[10pt]
    \label{eq:28}
    \nu_Q &= \nu_{0\rightarrow 1}
    - J'(J'+1)(B_1 - B_0)
    \\[10pt]
    \label{eq:29}
    \nu_R &= \nu_{0\rightarrow 1}
    + (B_1 + B_0)(J''+1) + (B_1 - B_0)(J''+1)^2
    \\[10pt]
    \label{eq:30}
    \nu_S &= \nu_{0\rightarrow 1} + B_1(J''+2)(J''+3) - B_0J''(J''+2)
\end{align}

Where $\nu_{0\rightarrow 1}$ represents the band origin, $B_0$ and $B_1$ represent the rotational constants for their appropriate vibrational levels. As previously mentioned, for our code we use the solution calculated by TOSH (equation \ref{eq:8}) as the band origin.

Here is a reminder of the labelling for the line series:
$\Delta J = -2$ (O Branch), $\Delta J = -1$ (P Branch), $\Delta J$ = 0 (Q Branch), $\Delta J = +1$ (R Branch) and $\Delta J = +2$ (S Branch).

Finally, the code combines the intensity calculations with the transition positions to produce the synthetic spectra.

\subsection{Prometheus}
\label{subsection:Prometheus}
Prometheus is the name given to our code, which is written in Python. It is maintained by the authors and is available on the Milne Centre Github (\url{https://github.com/Milne-Centre/Prometheus}). The version of Prometheus used to obtain the results of this paper is archived at  \url{https://doi.org/10.5281/zenodo.5494420}. 

Our code is designed to be an approximate method of simulating spectra, potentially trading off accuracy for simplicity and speed. The main idea behind this code is to offer a new complementary approach to modelling the vast amount of molecules which have astrophysical and astrobiological importance.

Our analysis may have applications beyond the SBP catalogue, as we have done ourselves within this work by modelling carbon monoxide.

Prometheus primarily implements the TOSH theory detailed in section \ref{subsection:TOSH} and makes use of the shift parameter, $\sigma$ to calculate band origins and provide anharmonic corrections to spectroscopic constants. To our knowledge this is the first time it has been used to provide anharmonic corrections to rotational constants, rather than purely being used for anharmonically corrected vibrational frequency calculations.

Prometheus then produces a stick spectrum (without any line broadening effects) using the constants it has determined.

\section{Results and Discussion}
\label{section:Results}
\label{subsection:comparison_catalogues}
The spectra Prometheus produces can be compared to different sources to evaluate the accuracy of our method. This doesn’t necessarily mean that if the transitions line up well with another spectrum that Prometheus is “correct”, rather that it has a better agreement with one of the mainstream methods.

When producing spectra, our program can either use purely TOSH, purely harmonic or a mixture of both sets of constants (such as the hybrid approach which we use within this study, detailed further in section \ref{subsection:rot_constants}). Additionally, Prometheus is capable of using any other experimental/literature values for the spectroscopic constants. A caveat however is that although the experimental/literature results may be more accurate, they are still bound to our code's capabilities and don't include all the effects that ExoMol or HITRAN have, for example.

Within this paper we will be only considering the data from HITRAN and ExoMol, and not RASCALL. This is because, as previously mentioned, the spectra produced by RASCALL 1.0 focus mainly on approximate band centres \citep{19SoPeSe} rather than rotational structure.

The comparison data we use to evaluate the performance of Prometheus is briefly described below:
\begin{itemize}
    \item Literature: We used the spectroscopic constants published in papers that are well known and commonly cited in the spectroscopic community. We have included these results to highlight the theoretical ``best" that our code could produce without a TOSH approximation. For the discussion of each diatomic, the appropriate paper has been stated and annotated on the results. 
    \item HITRAN2016: data from HITRAN’s 2016 release. HITRAN is an acronym for high-resolution transmission molecular absorption database and is a compilation of spectroscopic parameters that a variety of computer codes use to predict and simulate the transmission and emission of light in the atmosphere \citep[][]{17GoRoHi}. Data was obtained from the \url{https://hitran.org} website, using the linelist data access interface, and directly read into our code. This data is currently acting as the co-optimal result along with ExoMol. We normalised the intensities of the lines, to match the relative intensities produced by Prometheus. Future work will potentially look at the intensities of the rovibrational spectrum - whether that be regarding opacity functions or using units of atmospheric concentration rather than the current relative intensities.
    \item ExoMol: A database which provides high-accurate and complete linelists for application in hot astrophysical environments \citep[][]{20TeYuAl}. The ExoMol data structure can be used to generate lifetimes, cooling functions and partition functions.  More details can be found in \citet{16TeYuAl} and \citet{21ChRoYu}.
    The website: \url{http://www.exomol.com/} is the source of the ExoMol data used for comparisons. Like the HITRAN we have normalised the intensities of the lines, to allow comparisons in intensities.
    \item Harmonic: This is simply using Prometheus constants, calculated from the potential energy curve (PEC), without any anharmonic corrections. For example, the band origin is the harmonic frequency, $\omega$, instead of using equation \ref{eq:7} for the origin. This data is meant to highlight how harmonic methods, whilst simple and easy to do, is typically the weakest in accuracy and therefore there is a need for anharmonic corrections. 
\end{itemize}

The comparison spectra that are available (ExoMol or HITRAN or both) have been inverted on the figures below. This is merely an aesthetic choice to prevent the figures from becoming difficult to interpret, by having up to 4 sets of data layered upon each other.


We will briefly outline the colour scheme used throughout the spectra. Prometheus is designated as red, and the spectra produced by literature constants via Prometheus as blue. HITRAN 2016 spectra is denoted by the colour black whereas ExoMol spectra is the colour purple. Finally the harmonic spectra is green. 

The computed error on the line positions for the spectra generate by Prometheus are typically less than 0.1 wavenumbers, except H$_2$ which is less than 4 wavenumbers. We did not include error bars on each line position in the figures for clarity.


\begin{table*}
\begin{center}
\begin{tabular}{|c|c|cc|cc|cc|} 
\cline{3-8}
\multicolumn{1}{c}{} 
&            
&\multicolumn{2}{c|}{\begin{tabular}[c]{@{}c@{}}Band Origin\\ $\nu_{0\rightarrow 1}$ (cm$^{-1}$)\end{tabular}}     
&\multicolumn{2}{c|}{\begin{tabular}[c]{@{}c@{}}Rotational Constant\\ $B_0$ (cm$^{-1}$)~\end{tabular}}           
&\multicolumn{2}{c|}{\begin{tabular}[c]{@{}c@{}}Rotational Constant\\ $B_1$ (cm$^{-1}$)\end{tabular}}  
\\ 
\cline{3-8}
\multicolumn{1}{c}{} 
&            
&Calculated 
&$\Delta$Lit.
&Calculated
&$\Delta$Lit.
&Calculated 
&$\Delta$Lit.                             
\\ 
\hline
H$_2$             
& \citet{12CaKaPa}       
& 4161.17    
& ---          
& 59.3329  
& ---           
& 56.3732         
& ---                                      
\\
& Harmonic   
& $4411.45 \pm 0.13$                               
& $+250.28$ 
& \colorbox{green}{60.9513}                             
& $+1.6184$ 
& $60.9513$                                             
& $+4.5781$ 
\\
& TOSH 
& \colorbox{green}{4174.71}$\pm 2.53$      
& $+13.54$ 
& $54.2350 \pm 0.0015$          
& $-5.0979$ 
& \colorbox{green}{54.2350}$\pm 0.0015$          
& $-2.1382$ 
\\
\hline
O$_2$
& \citet{14YuDrMi}       
& $1556.39$ 
& ---                     
& $1.4377$ 
& ---                             
& $1.4219$ 
& ---                       
\\
& Harmonic   
& 1580.00                    
& $+23.61$ 
& \colorbox{green}{1.4456}                      
& $+0.0079$ 
& 1.4456                                 
& $+0.0237$ 
\\
& TOSH 
& \colorbox{green}{1556.43}$\pm 0.07$   
& $+0.04$ 
& 1.4257      
& $-0.0120$ 
& \colorbox{green}{1.4257} 
& $+0.0038$ 
\\
\hline
N$_2$             
& \citet{00BeRaXX}       
& $2329.91$         
& ---                  
& 1.9896             
& ---        
& 1.9722           
& ---                                        
\\
& Harmonic   
& $2359.56 \pm 0.02$           
& $+29.65$ 

& \colorbox{green}{1.9982}       
& $+0.0086$ 
& 1.9982 
& $+0.0260$ 
\\
& TOSH 
& \colorbox{green}{2329.85}$\pm 0.14$ 
& $-0.06$ 
& 1.9753         
& $-0.0143$ 

& \colorbox{green}{1.9753}    
& $+0.0031$ 
\\
\hline
CO                   
& \citet{04CoHaXX}       
& 2143.27      
& ---               
& 1.9225         
& ---                
& 1.9050           
& ---                                       
\\
& Harmonic   
& $2173.07 \pm 0.02$                
& $+29.80$ 
& \colorbox{green}{1.9316}                 
& $+0.0091$ 
& 1.9316                               
& $+0.0266$ 
\\
& TOSH 
& \colorbox{green}{2143.15}$\pm 0.08$ 
& $-0.12$ 
& 1.9080          
& $-0.0145$ 
& \colorbox{green}{1.9080}          
& $+0.0030$ 
\\
\hline
\end{tabular}
\caption{A comparison of the spectroscopic constants between Prometheus (TOSH), the literature and the harmonic method. $\Delta$Lit. represents the difference between the calculated value and the appropriate literature. The highlighted values represent the values which Prometheus uses to create spectra, with the band origin and upper rotational coming from TOSH theory and the ground level rotational constant coming from the harmonic approximation. The band origins are calculated to 2 decimal places. The rotational constants have been calculated to 4 decimal places. Errors have been included where available (some literature sources did not provide them) and other errors have been omitted if they were of a lower order than the rounding criterion of the data. }
\label{table:comparing_spec_values}
\end{center}
\end{table*}

\subsection{Molecular Hydrogen}
\label{subsection:ResultsH2}

The first biosignature molecule is molecular hydrogen. H$_2$ is the most abundant molecule in the Universe by orders of magnitude \citep[][]{17WaBrCa}. It is difficult to observe directly in the interstellar medium due to a lack of permanent electric dipole moment and most transitions are of quadrupolar nature (and extremely weak) \citep[][]{49XXXX, 19RoAbCz}, or in emission for selected environments (see \citet[][]{04HaWaVe} for a review). Indirect techniques, such as the amount of dust present using the gas-to-dust ratio (GDR) \citep[][]{17JoSeMi}, also allow some degree of detection. See \citet[][]{20SeHuPe} for further discussion of possible H$_2$ presence and detection on exoplanets. From a biology standpoint, many microorganisms produce hydrogen as primary product through metabolism under anaerobic conditions \citep[][]{73ScXXXX}. 

For H$_2$, data from HITRAN2016 \citep[][]{17GoRoHi} was available \citep[specifically, the line list comes from][]{11KoPiLa, 98WoSiDa}. The spectroscopic constants were obtained from \citet{12CaKaPa}. Additionally, spectra was available from ExoMol \citep[][]{19RoAbCz}.

In Table \ref{table:comparing_spec_values}, \citet{12CaKaPa} determined the band origin to be $4161.17~\rm{cm}^{-1}$. The harmonic approach gives $4411.45~\rm{cm}^{-1}$ which deviates from the literature by $250.28~\rm{cm}^{-1}$. TOSH, with a result of $4174.71~\rm{cm}^{-1}$, manages to produce a band origin with a difference of only $13.54~\rm{cm}^{-1}$ --- an order of magnitude less than the harmonic approach. 

The experimental ground state rotational constant, $B_0$, was determined by \citet{12CaKaPa} to be $59.3329~\rm{cm}^{-1}$. The harmonic method approximates $B_0$ better, with a value of $60.9513~\rm{cm}^{-1}$, than TOSH which appears to drastically over-correct, leading to a value of $54.2350~\rm{cm}^{-1}$. 

However, since the predicted rotational constants has no vibrational dependence for TOSH or the harmonic model, the reverse is true for $B_1$. \citet{12CaKaPa} determined $56.3732~\rm{cm}^{-1}$, with TOSH and harmonic now differing by $2.1382~\rm{cm}^{-1}$ and $4.5781~\rm{cm}^{-1}$, respectively. 

\begin{figure}[H]
    \centering
    \includegraphics[width=\columnwidth]{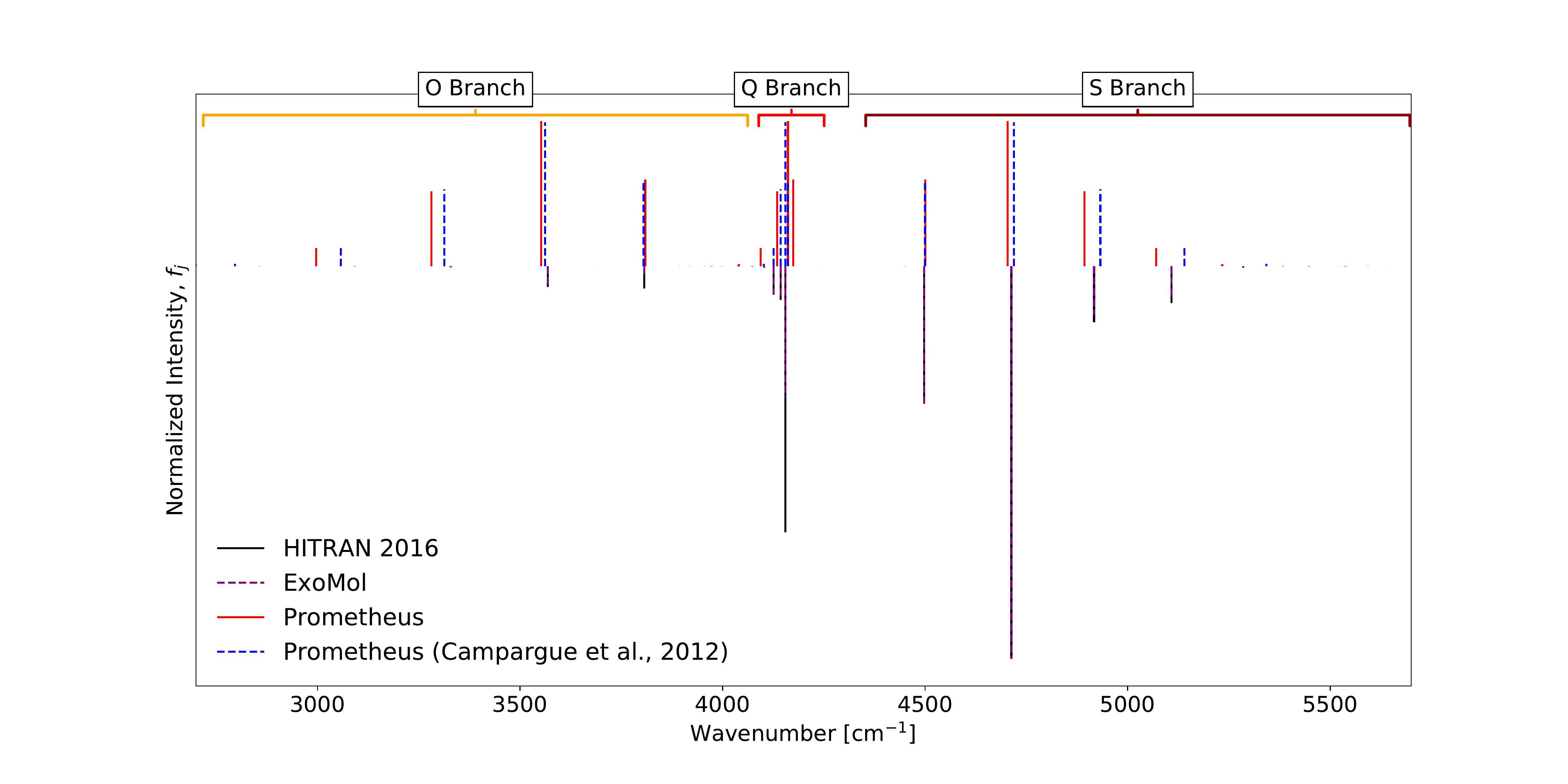}
    \caption{A comparison of the H$_2$ spectra produced by Prometheus, ExoMol \citep[][]{19RoAbCz} and Prometheus using the spectroscopic constants from \citet{12CaKaPa}. The intensities for all sets of data have been normalised and therefore are relative values. Note that HITRAN and ExoMol transitions overlap with each other.}
    \label{fig:H2/H2_Rovibe_Spectrum_Fundamental_PROvNISTvHITRAN2016vEXOMOL}
\end{figure}

\begin{figure}[H]
    \centering
    \includegraphics[width=\columnwidth]{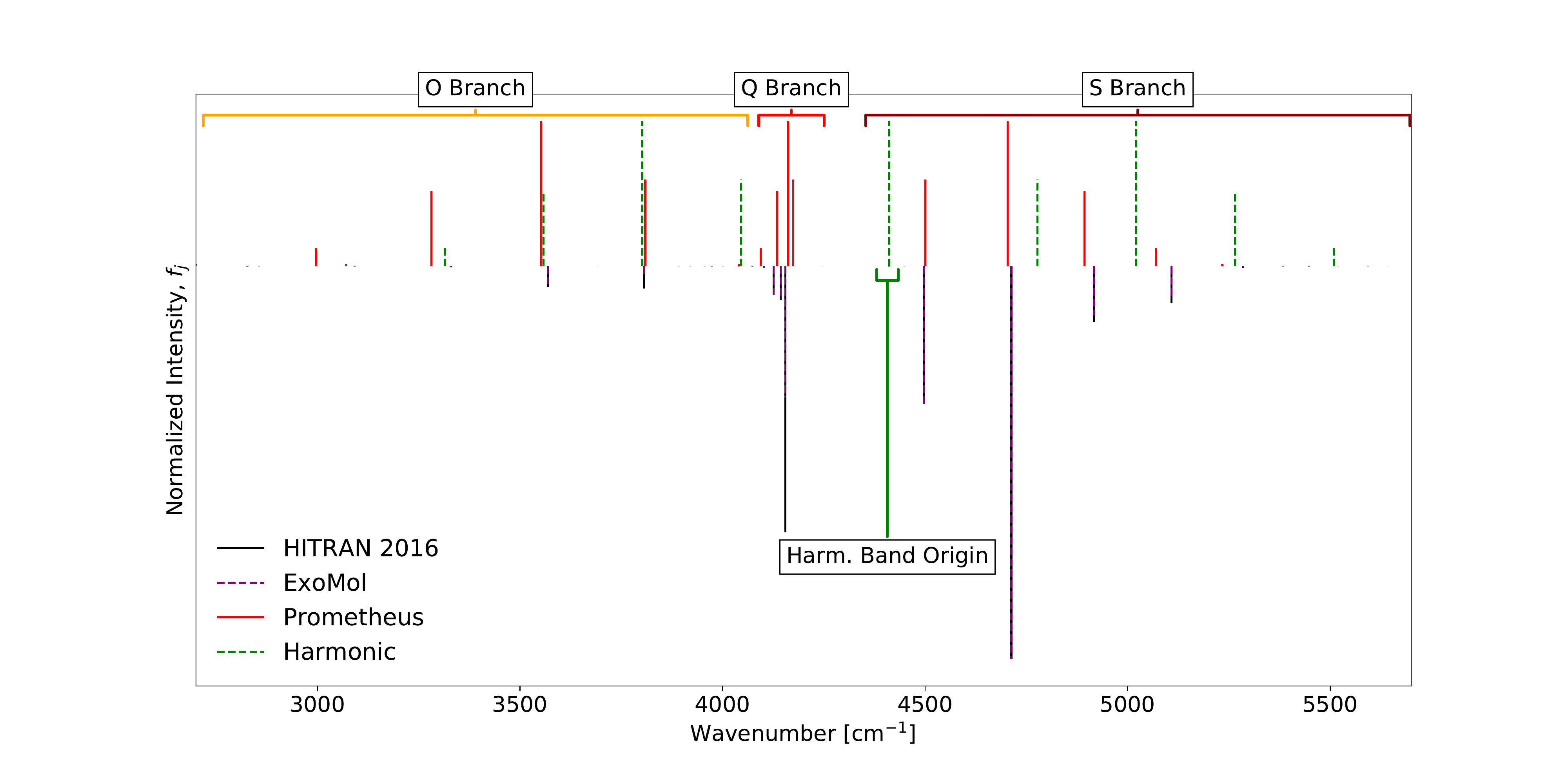}
    \caption{A comparison of the H$_2$ spectra produced by Prometheus, ExoMol \citep[][]{19RoAbCz} and the Harmonic method. The intensities for all sets of data have been normalised and therefore are relative values. Note that HITRAN and ExoMol transitions overlap with each other.}
    \label{fig:H2/H2_Rovibe_Spectrum_Fundamental_PROvHARMvHITRAN2016vEXOMOL}
\end{figure}

Figures \ref{fig:H2/H2_Rovibe_Spectrum_Fundamental_PROvNISTvHITRAN2016vEXOMOL} and \ref{fig:H2/H2_Rovibe_Spectrum_Fundamental_PROvHARMvHITRAN2016vEXOMOL}, show the resulting simulated spectra from Prometheus compared to ExoMol, data by \citet{12CaKaPa} and the harmonic method.

Two major effects need to be considered when the rovibrational spectrum of H$_2$ is analysed. First, this homonuclear molecule exhibits a center of symmetry. As a consequence of this, it is Raman active and IR inactive with O, Q and S branches present in its spectra \citep{12CaKaPa}. Prometheus will model a molecule in the correct region by considering symmetry of the diatomic using the initial mass inputs.

As also mentioned by \citet{12CaKaPa}, a molecule composed of nuclei with non-zero nuclear spin (such as H$_2$) will exhibit spectral lines that show an alternation in intensity. Unfortunately Prometheus does not currently model this. Instead the intensities have the typical Boltzmann distribution regardless of the nuclear spin of the molecule's nuclei components.

A key aspect of Figure \ref{fig:H2/H2_Rovibe_Spectrum_Fundamental_PROvNISTvHITRAN2016vEXOMOL}, is how well all three band origins align. Using spectroscopic constants of \citet{12CaKaPa}, Prometheus can quite accurately reproduce the ExoMol positions for the transitions. Prometheus TOSH approach however appears to slightly overestimate the band origin by about 20~cm$^{-1}$. This is likely due to the strong quantum nature of the hydrogen nuclei and thus this molecule displays strong anharmonic effects that are less well modeled by the TOSH approach.
Regardless, Figure \ref{fig:H2/H2_Rovibe_Spectrum_Fundamental_PROvNISTvHITRAN2016vEXOMOL}  shows all three approaches are in relatively good agreement with one another.

Regarding the Q-branch, Prometheus displays a larger spread of transitions than ExoMol and \citet{12CaKaPa}. The Prometheus central transitions potentially would roughly line up provided the band origin was corrected. Once again \citet{12CaKaPa} adequately models the ExoMol data. It does falter slightly with the distribution of the branch, as the \citet{12CaKaPa} Q branch appears to model a greater spread in the transitions than ExoMol. 

The O branch, indicated by the labels on Figures \ref{fig:H2/H2_Rovibe_Spectrum_Fundamental_PROvNISTvHITRAN2016vEXOMOL} and \ref{fig:H2/H2_Rovibe_Spectrum_Fundamental_PROvHARMvHITRAN2016vEXOMOL}, shows an interesting result. \citet{12CaKaPa} appears to decrease slightly in accuracy the higher the transition, whereas Prometheus does the same but the decrease in accuracy is larger. This deterioration is to such a point it becomes difficult to match the ExoMol's transitions to Prometheus.

For the S branch (also indicated by the label on Figures \ref{fig:H2/H2_Rovibe_Spectrum_Fundamental_PROvNISTvHITRAN2016vEXOMOL} and \ref{fig:H2/H2_Rovibe_Spectrum_Fundamental_PROvHARMvHITRAN2016vEXOMOL}), Prometheus seems to initially fare quite well. The Prometheus transition positions appear to slightly increase in difference to the ExoMol data with each higher transition. Possibly if the origins had aligned better Prometheus would be able to continue to quite effectively model this branch at higher rotational transitions. As before, \citet{12CaKaPa} gives a satisfactory level of comparison to ExoMol, until the higher transitions are reached at which point it begins to falter. Some difficulty occurs with the comparisons at higher transitions as the relative intensities for the ExoMol data are very low.
The ExoMol S branch seems to have a far greater weighting than the O branch, something which is not modelled in any other spectra.

The difficulties in describing the higher rotational transitions can be partly explained by the fact that hydrogen is a light molecule and thus distortion constants are required to accurately model anything beyond $J = 0, 1$. For example, it was found by \citet{83JeBrXX} that to accurately fit up to $J = 5$, four rotational constants (B, D, H and L) were required. This presumably explains Prometheus' inaccuracies for the hydrogen spectra, which are not as pronounced in the other heavier molecules considered within these simulations.

Finally, we compare the Prometheus spectrum with the Harmonic model in Figure \ref{fig:H2/H2_Rovibe_Spectrum_Fundamental_PROvHARMvHITRAN2016vEXOMOL}. As discussed earlier, it can be seen that Prometheus slightly overestimate the origin, whereas the harmonic method substantially overestimates it. Due to this, it is difficult to comment on much of the harmonic spectrum, as it barely matches ExoMol.

We can see that despite Prometheus also having some difficulties with modelling H$_2$, particularly at higher transitions and in the O branch, it still performs better than the Harmonic approach. Certainly, this shows anharmonic corrections are required to produce a qualitatively correct result.

\subsection{Molecular Oxygen}
\label{subsection:ResultsO2}
The second biosignature is molecular oxygen. Oxygen is the third most abundant element in the universe, yet molecular oxygen is one of the most elusive molecules \citep[][]{20WaLiGo}. This is often cited to be due to O$_2$'s high chemical reactivity and lack of electric dipole moment \citep[][]{18LuMoLu}. Even now, a comprehensive picture of oxygen chemistry in interstellar environments is still missing \citep[][]{20WaLiGo, 10WaSmHe, 06AGSeXX}.

Oxygen is crucial to our understanding of most life on Earth with oxygenic photosynthesis being the dominant producing metabolism on our planet \citep[][]{14DoSeCl}. Concerning exoplanets observations, oxygen is a potential biosignature since it can be produced through photosynthesis processes. However, it is also a possible false positive since it can be formed in larger quantities through abiotic processes such as runaway greenhouse effects \citep[][]{17MeXXXX}.

Oxygen's potential to act as a false positive biosignature is rooted in the diversity of exoplanets. For example, on Earth there are no abiotic processes that would produce it in large abundance \citet{18MeReAr}. It has been shown for a very different star and planetary system, O$_2$ could be generated in a planetary environment without life \citep[][]{14WoPiXX}.

The literature spectroscopic constants for O$_2$ were obtained from  \citet{14YuDrMi}, and our values are compared to data from the HITRAN 2016 release \citep[][]{17GoRoHi, 98GaGoRo, 14YuDrMi} only, as no ExoMol data was available. It is worth noting that the HITRAN data contains not Raman but magnetic dipole and electric quadrupole IR transitions, which may lead to some discrepancies between the data sets.

Like $\rm{H_2}$, $\rm{O_2}$ has a centre of symmetry which causes it to be Raman active but IR inactive. This once again means that the O, Q and S branches are present in the spectra. In addition the symmetry also means the effects of nuclear spin will be observed, but these effects differ to that of the other homonuclear diatomic molecules.

\begin{figure}[H]
    \centering
    \includegraphics[width=\columnwidth]{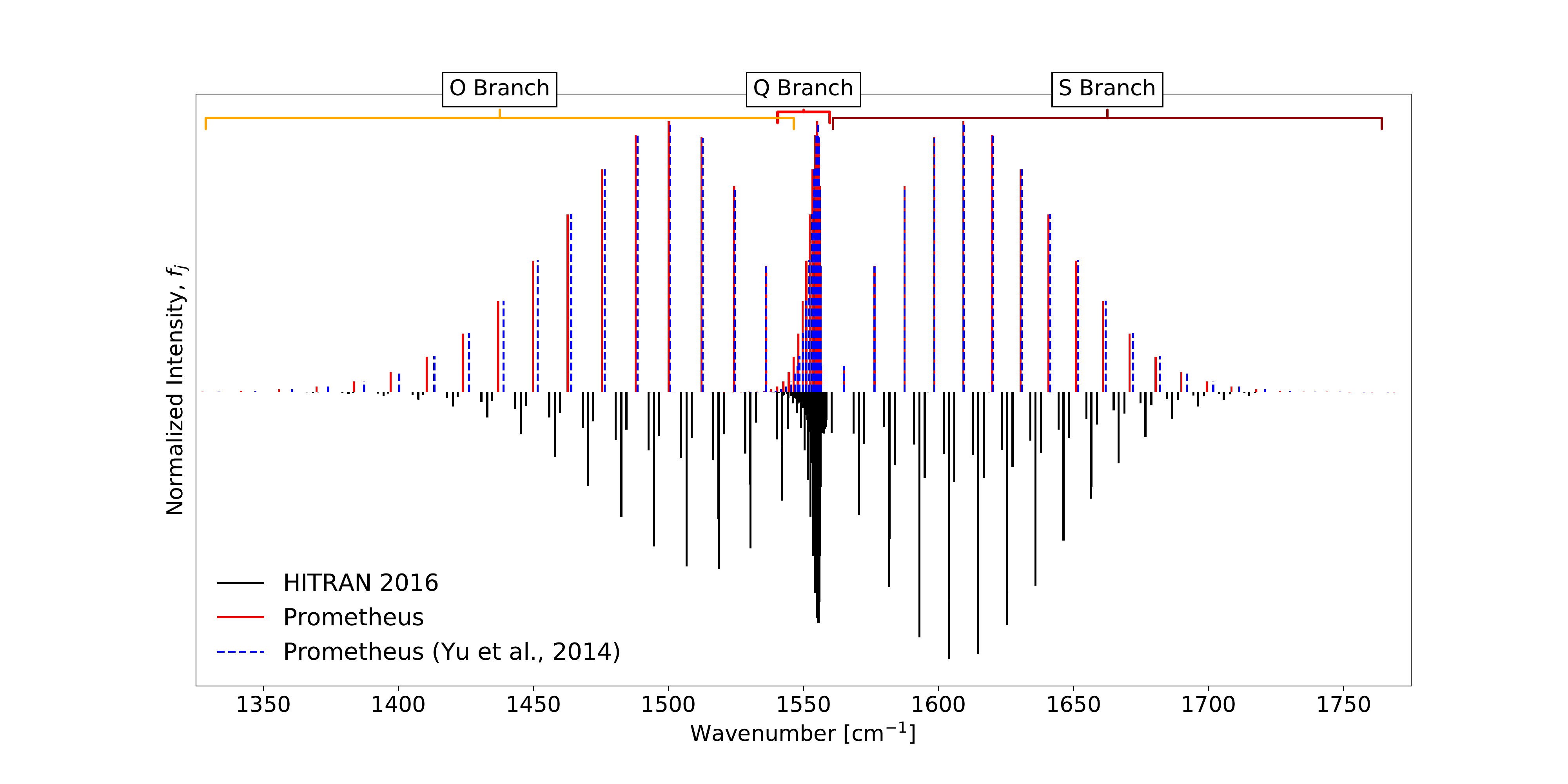}
    \caption{A comparison of the O$_2$ spectra produced by Prometheus, Prometheus using the spectroscopic constants from \citet{14YuDrMi} and HITRAN 2016 data \citep[][]{17GoRoHi, 98GaGoRo, 14YuDrMi}. The intensities for all sets of data have been normalised and therefore are relative values.}
    \label{fig:O2/O2_Rovibe_Spectrum_Fundamental_PROvNISTvHITRAN2016}
\end{figure}

Unlike the other diatomic molecules in this study, $\rm{O_2}$ has a zero nuclear spin which has a different effect on the spectra than the alternating intensities effect. Instead, every rotational level with an even value is absent from the spectra. This presents itself as a spectra with seemingly large gaps between the rotational transitions \citet{14YuDrMi} (See Fig.~\ref{fig:O2/O2_Rovibe_Spectrum_Fundamental_PROvNISTvHITRAN2016}).

$\rm{O_2}$ has a triplet sigma state and exhibits a triplet structure for each of the remaining transition lines, due to the splitting of the rotational levels. This is another effect that Prometheus does not currently model and needs to be taken into consideration when evaluating spectra. However, the weighting of the relative intensities due to triplet splitting have been included into Prometheus, to allow for ease of comparison.

\begin{figure}[H]
    \centering
    \includegraphics[width=\columnwidth]{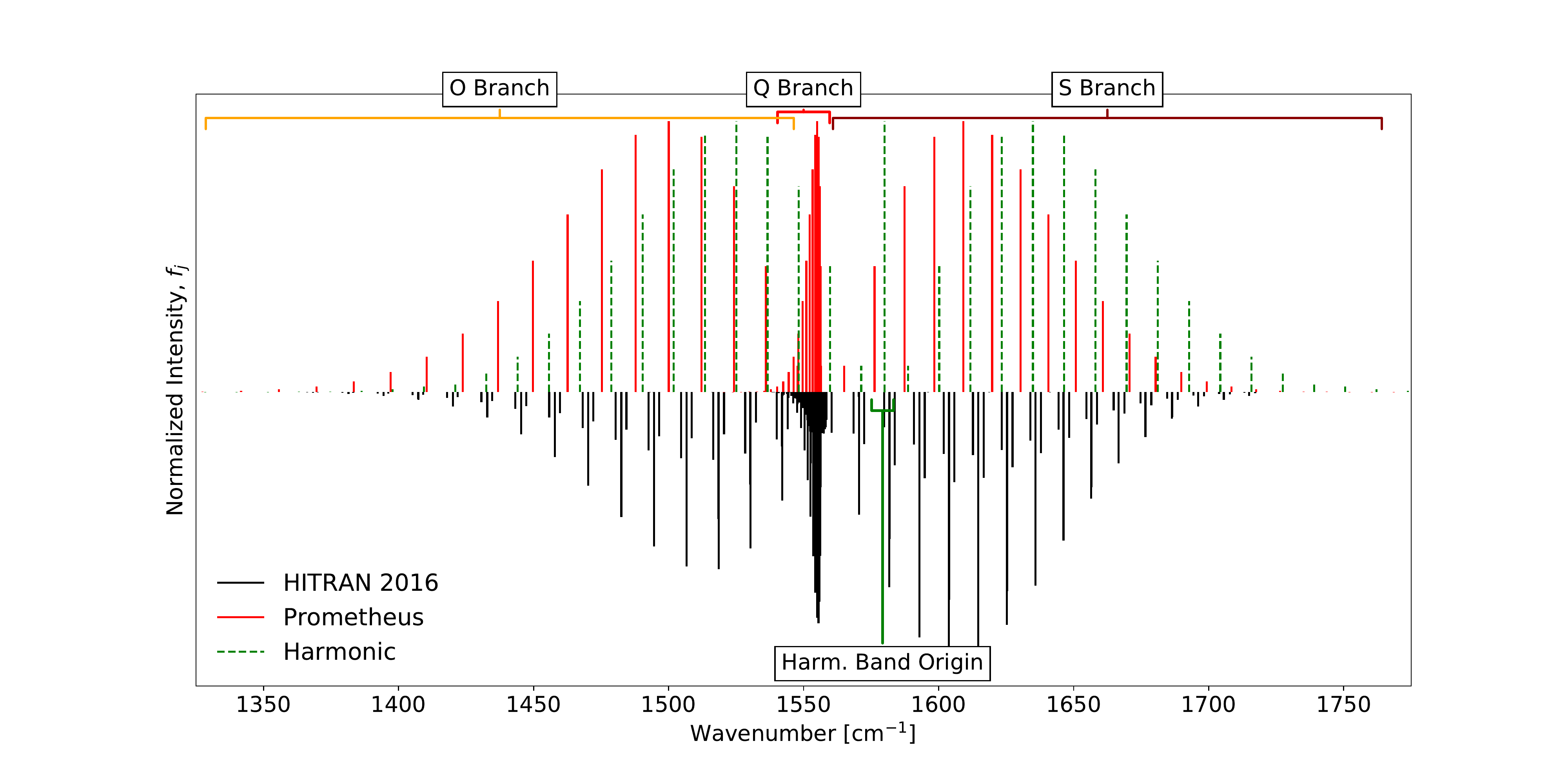}
    \caption{A comparison of the O$_2$ spectra produced by Prometheus, harmonic method and HITRAN 2016 data \citep[][]{17GoRoHi, 98GaGoRo, 14YuDrMi}. The intensities for all sets of data have been normalised and therefore are relative values.}
    \label{fig:O2/O2_Rovibe_Spectrum_Fundamental_PROvHARMvHITRAN2016}
\end{figure}

When considering the band origins (see Table \ref{table:comparing_spec_values}), TOSH produces a better approximation to the literature than the harmonic method, despite the points mentioned previously. \citet{14YuDrMi} reports the band origin to be at $1556.39~\rm{cm}^{-1}$, and this is extremely well estimated by TOSH, with a result of $1556.43~\rm{cm}^{-1}$. The harmonic approximation on the other hand displays a significant displacement, roughly 24~cm$^{-1}$. We also note again that the harmonic approximation provides a better approximation of $B_0$, with TOSH doing the same for $B_1$ .

In Figure \ref{fig:O2/O2_Rovibe_Spectrum_Fundamental_PROvHARMvHITRAN2016}, we can quite clearly see the shift in band origin has a significant effect on the harmonic approximations ability to mimic the literature.

Prometheus, by using our hybrid methodology, models the Q branch well \citep[in good agreement with][]{14YuDrMi}. This is something the purely harmonic method (or a purely TOSH approach) repeatedly fails at due to the fixed rotational constants between levels.

\subsection{Molecular Nitrogen}
\label{subsection:ResultsN2}
The third diatomic biosignature is molecular nitrogen. This is a key species in cosmology and has been observed in various galactic environments \citep[][]{18VaDvOl}. Although direct observation of N$_2$ is often difficult, as it lacks strong pure rotational or vibrational lines \citep[][]{13LiHeVi}. 

Within our own solar system we find N$_2$ present in various planetary and satellites' atmospheres; it accounts for 3.5\% \citep[][]{80OyCaWo} of the Venusian atmosphere, 78.1\% \citep[][]{02CoXXXX} of the terrestrial atmosphere, 2.8\%  \citep[][]{17FrTrMa} of the Martian atmosphere, and perhaps most significantly, 98.4\% \citep[][]{82StShXX} of Titan's atmosphere.

From a biogenic perspective, N$_2$ is an essential ingredient for the building blocks for life as we know it\citep[][]{18SpLaGr} since it is required, along with carbon and phosphorus, for the formation of nucleic acids and proteins \citep[][]{19LaSpGr}. 

The spectroscopic constants were obtained from 
\citet{00BeRaXX}. Our N$_2$ results are compared to the harmonic approximation, literature optimal results and the data from the HITRAN 2016 release \citep[][]{17GoRoHi, 07LiLeXX, 06LeHuJa}. The available ExoMol data was not suitable for our study as it provides the \citet[][]{18WeCaCr} data, which focuses on a different region of the spectrum ($4500$-$11,000~\textrm{cm}^{-1}$), rather than the ground state. 

Finally $\rm{N_2}$, like $\rm{O_2}$  and $\rm{H_2}$, has a center of symmetry and thus the spectrum is Raman active but IR inactive. As $\rm{N_2}$ is composed of nuclei of non-zero nuclear spin, the HITRAN spectrum shows the characteristic alternation in intensities (again, not modelled here by our Prometheus approach).

By looking at the results of Table \ref{table:comparing_spec_values}, we see that N$_2$ rotational constants follow the typical pattern of the previous molecules within this study. \citet{00BeRaXX} determined the first excited level rotational constant to be $1.9722~\rm{cm}^{-1}$ and the ground rotational constants to be $1.9896~\rm{cm}^{-1}$. The TOSH approach, with a fixed value of $1.9753~\rm{cm}^{-1}$, over-estimates the ground state but provides a good approximation for the first excited level. Whereas the harmonic approximation, with a constant value of $1.9982~\rm{cm}^{-1}$, achieves the opposite.

\begin{figure}[H]
    \centering
    \includegraphics[width=\columnwidth]{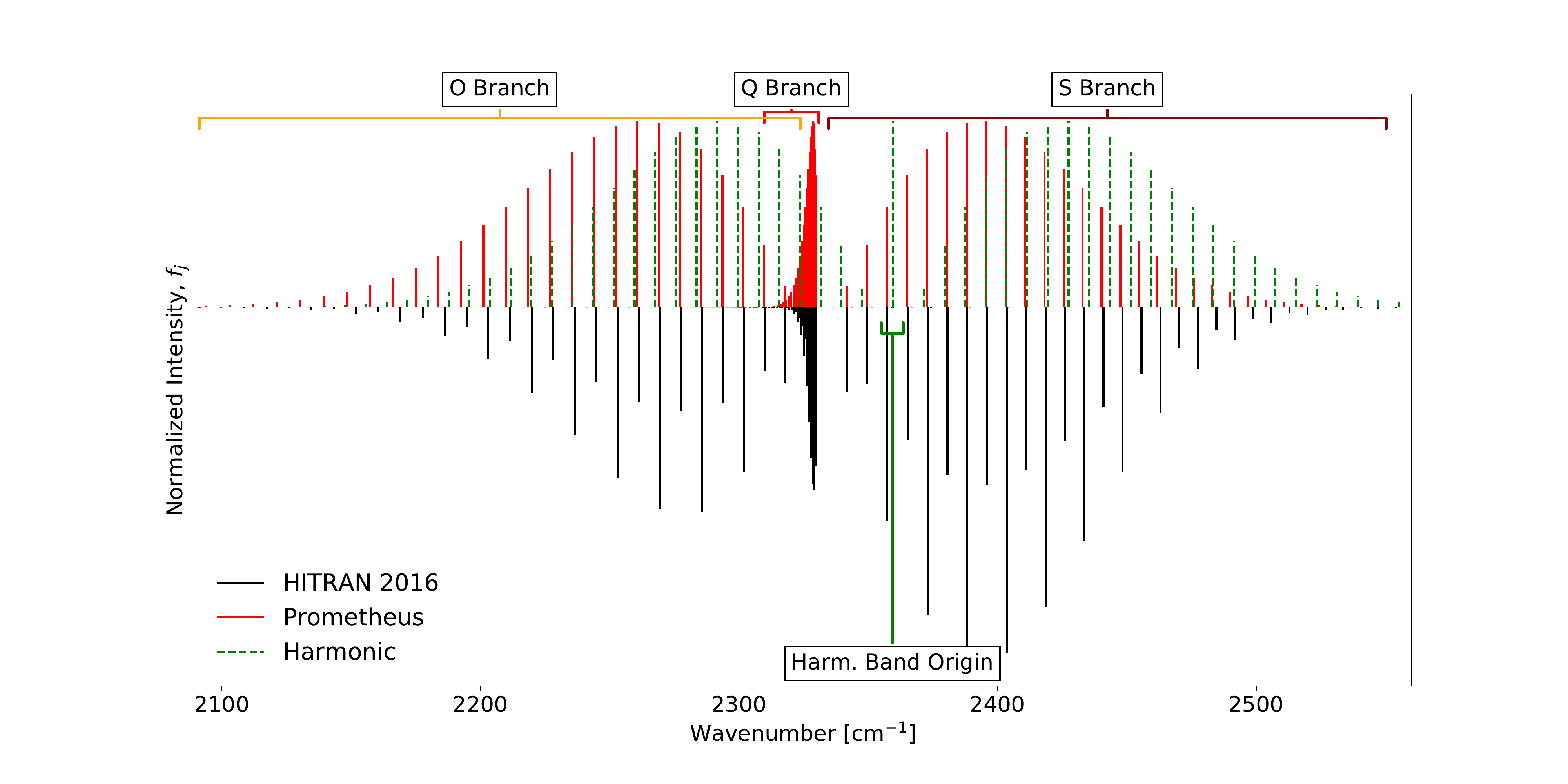}
    \caption{A comparison of the N$_2$ spectra produced by Prometheus,  the harmonic method and HITRAN 2016 \citep[][]{17GoRoHi, 07LiLeXX, 06LeHuJa} data. The intensities for all sets of data have been normalised and therefore are relative values.}
    \label{fig:N2/N2_Rovibe_Spectrum_Fundamental_PROvHARMvHITRAN2016}
\end{figure}

We can see in Table \ref{table:comparing_spec_values} that for N$_2$, TOSH ($2329.85~\rm{cm}^{-1}$) once again provides a superior approximation of the band origin than the harmonic approximation ($2359.56~\rm{cm}^{-1}$), when comparing to the literature value ($2329.91~\rm{cm}^{-1}$). In Figure \ref{fig:N2/N2_Rovibe_Spectrum_Fundamental_PROvHARMvHITRAN2016}, we can draw comparisons between HITRAN, Prometheus and the \citet{00BeRaXX} constants used in a Prometheus spectrum. As expected from using the TOSH band origin, Prometheus provides an excellent approximation of both \citet{00BeRaXX} and HITRAN.

For the lower rotational transitions of both the O and S branches, Prometheus does a satisfactory job of the modelling the positions. In most cases it is lining up incredibly well with not only \citet{00BeRaXX} but also HITRAN data. As was the case for the other molecules this accuracy is lower for higher $J$ transitions.

Prometheus does an excellent job at approximating the Q branch for N$_2$ and the relative spread of the transitions of this branch is comparable to the HITRAN data.

For each increasing $J$ transition the position is incrementally over-estimated, but not to the point where the Prometheus spectrum is incorrectly matching lower number transitions to higher HITRAN ones. With this is mind, Prometheus is a very good approximation of both \citet{00BeRaXX} and HITRAN.

\begin{figure}[H]
    \centering
    \includegraphics[width=\columnwidth]{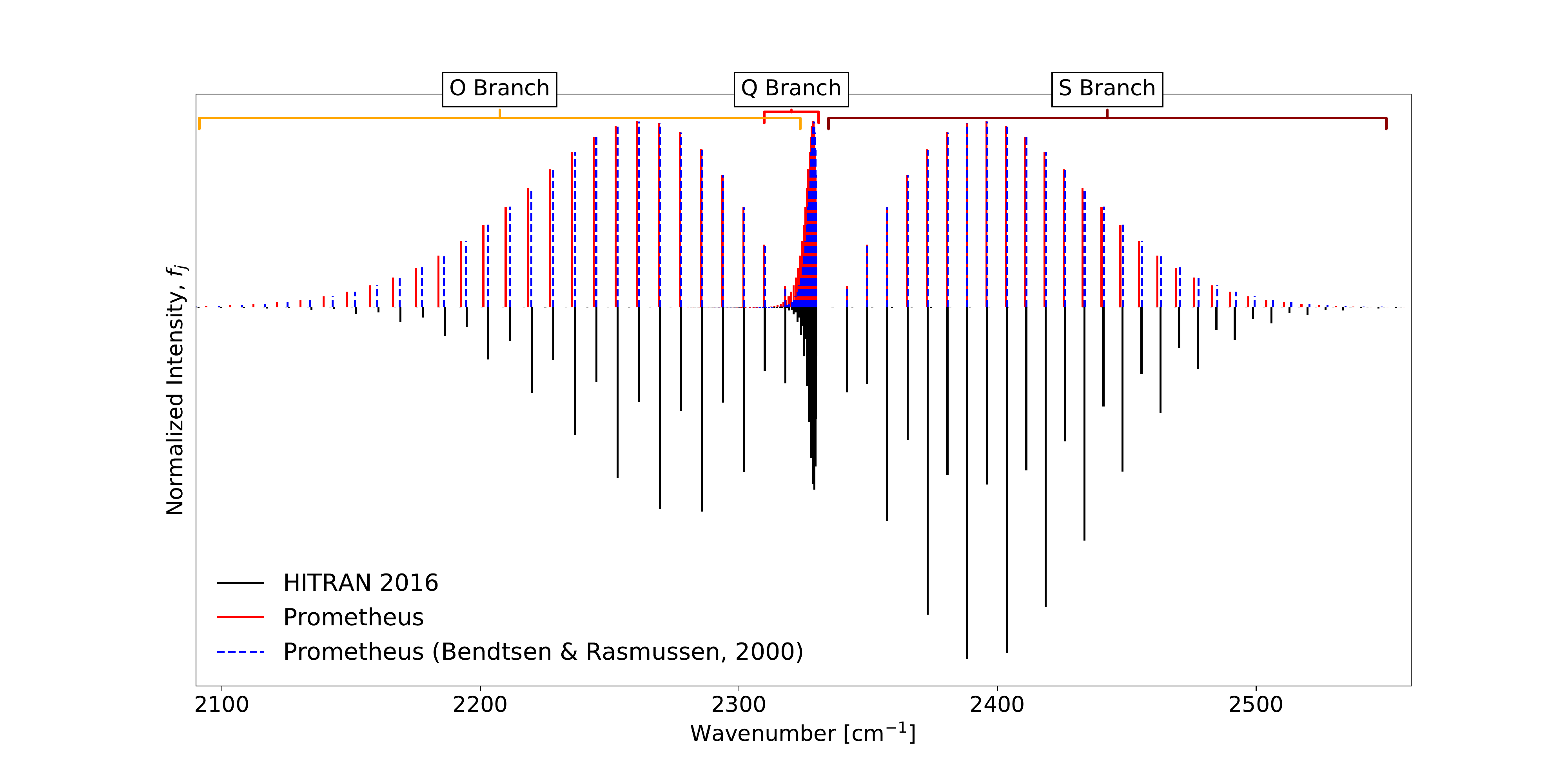}
    \caption{A comparison of the N$_2$ spectra produced by Prometheus, Prometheus using the spectroscopic constants from \citet{00BeRaXX} and HITRAN 2016 \citep[][]{17GoRoHi, 07LiLeXX, 06LeHuJa} data. The intensities for all sets of data have been normalised and therefore are relative values.}
    \label{fig:N2/N2_Rovibe_Spectrum_Fundamental_PROvNISTvHITRAN2016}
\end{figure}

A comparison with the harmonic spectrum is shown in Figure \ref{fig:N2/N2_Rovibe_Spectrum_Fundamental_PROvHARMvHITRAN2016}. The harmonic band origin is displaced from the literature by roughly 30 wave numbers (see Table \ref{table:comparing_spec_values}). This effectively renders any comparisons of the transition positions to the other spectra pointless. 

As was the case for H$_2$ the relative intensities of N$_2$ alternate in weighting, an effect that Prometheus does not currently model, once again causing the difficulties in drawing comparisons. The general intensities of the harmonic and the \citet{00BeRaXX} spectra have the same problem, which is expected, as it is produced via our Prometheus methodology. 

Comparing the relative intensities of Prometheus and harmonic to HITRAN does raise interesting points. The ``peak" for the O branch for HITRAN appears to occur 10-20 wavenumbers away from Prometheus. Potentially the alternating intensities is artificially causing the ``peak" to occur at a higher wavenumbers. The S branch does not appear to be as exaggerated. As discussed earlier, this discrepancy is not a major concern as intensities calculations are the not the focus of our present study.

We may conclude that, for $\rm{N_2}$, Prometheus is producing a better approximate spectrum than the harmonic method. Like the other molecules seen in previous sections, Prometheus' predictive power is reduced at higher rotational transitions. Presumably the uncorrected harmonic value for ground rotational constant and the TOSH approximate values for the rotational constant of the first excited level is having an effect on the later transitions positions.

\subsection{Carbon Monoxide}
\label{subsection:ResultsCO}

The last molecule considered in this work is carbon monoxide. Unlike the previous molecules, CO has a permanent dipole moment and therefore readily absorbs, even at the low temperatures found in molecular clouds \cite[][]{18WhJaXX}. More relevantly, CO has already been detected in an exoplanet's atmosphere \citep[HD 189733,][]{13KoBrSn}.

Within a biogenic context, CO is often classified as an anti-biosignature.  For example, for an inhabited planet, it could be difficult for CO to accumulate in the atmosphere as it acts as an energy source for some microbes on Earth \citep[][]{16WaTiLi}. Anthropologically, it can be formed instead when any organic substance is combusted incompletely and therefore can be an indication of not just life but intelligent life \citep[][]{00HoXXXX}.

The literature spectroscopic constants were obtained from \citet{04CoHaXX}. Our results for CO are compared to the harmonic approximation, optimal results and data from both HITRAN 2016 release \citep[][]{17GoRoHi, 04CoHaXX, 18DeBeSu, 15LiGoRo} and ExoMol's Li2015 line list \citep[][]{15LiGoRo}.

Carbon monoxide does not have a center symmetry therefore it is infrared (IR) active and Raman inactive. It is easily modelled by Prometheus to an accuracy akin to the HITRAN data at the lower transitions. Unlike the other biogenic diatomic molecules in this analysis, carbon monoxide solely demonstrates P and R branches, with no Q branch transitions.

\citet{04CoHaXX}, as shown in Table \ref{table:comparing_spec_values}, provided $1.9225~\rm{cm}^{-1}$ for $B_0$ and $1.9050~\rm{cm}^{-1}$ for $B_1$. As before, the harmonic approximation for $B_0$ is better ($1.9316~\rm{cm}^{-1}$) than TOSH's $1.9080~\rm{cm}^{-1}$. For the upper rotational constant, $B_1$, TOSH provides a rotational constant closer to literature.

\begin{figure}[H]
    \centering
    \includegraphics[width=\columnwidth]{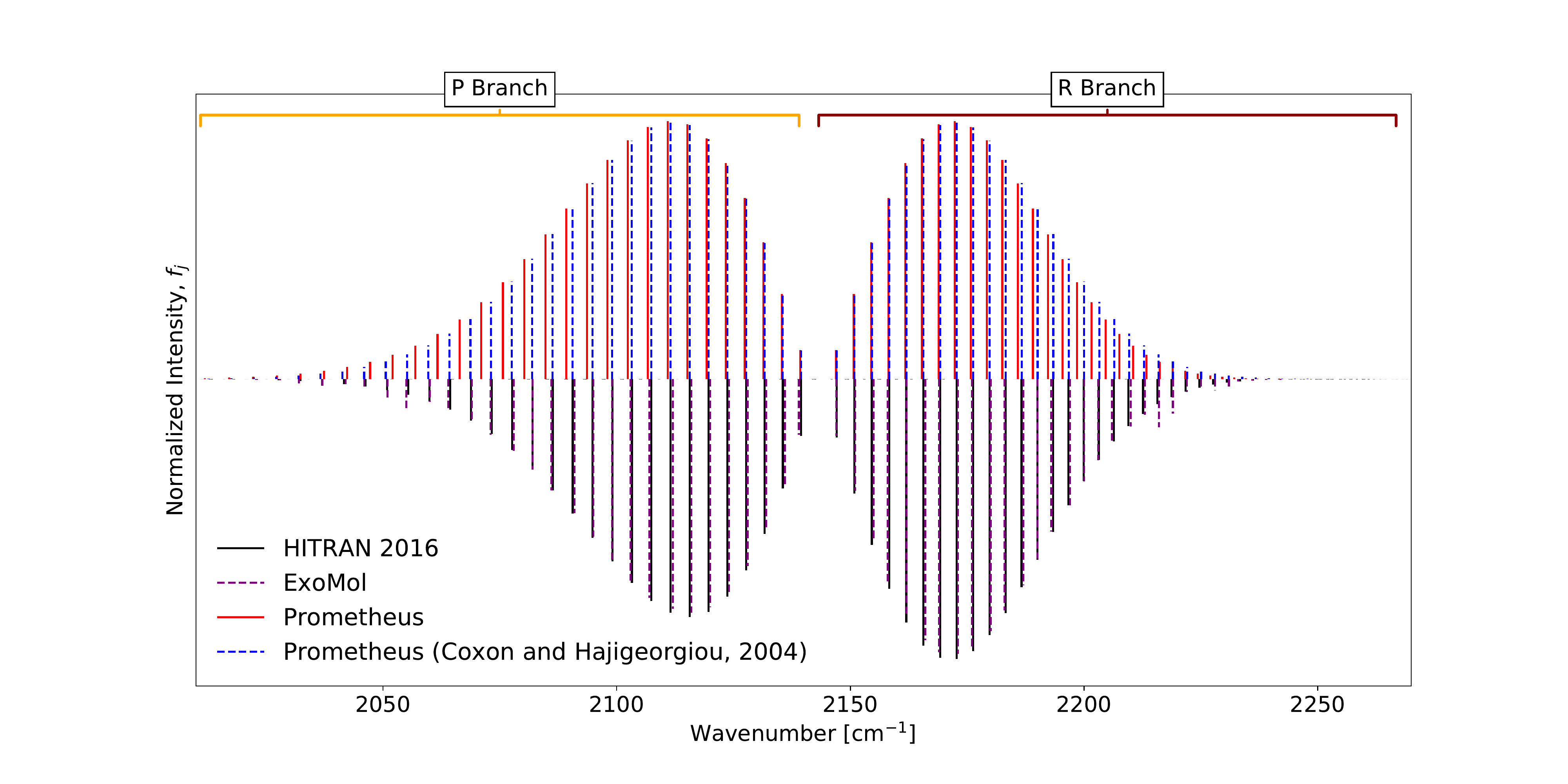}
    \caption{A comparison of the CO spectra produced by Prometheus, Prometheus using the spectroscopic constants from \citet{04CoHaXX}, ExoMol \citep[][]{15LiGoRo} and HITRAN 2016 data \citep[][]{17GoRoHi, 04CoHaXX, 18DeBeSu, 15LiGoRo}. The intensities for all sets of data have been normalised and therefore are relative values.}
    \label{fig:CO/CO_Rovibe_Spectrum_Fundamental_PROvNISTvHITRAN2016vExoMol_with_errs}
\end{figure}

The band origin of the Prometheus spectrum (using TOSH), is $2143.15~\rm{cm}^{-1}$, and easily lies within a single wavenumber of \citet{04CoHaXX} (at $2143.27~\rm{cm}^{-1}$). Figure \ref{fig:CO/CO_Rovibe_Spectrum_Fundamental_PROvNISTvHITRAN2016vExoMol_with_errs} shows that, even when compared to the HITRAN/ExoMol spectra, Prometheus' band origin still lies within a few wavenumbers. 

\begin{figure}
    \centering
    \includegraphics[width=\columnwidth]{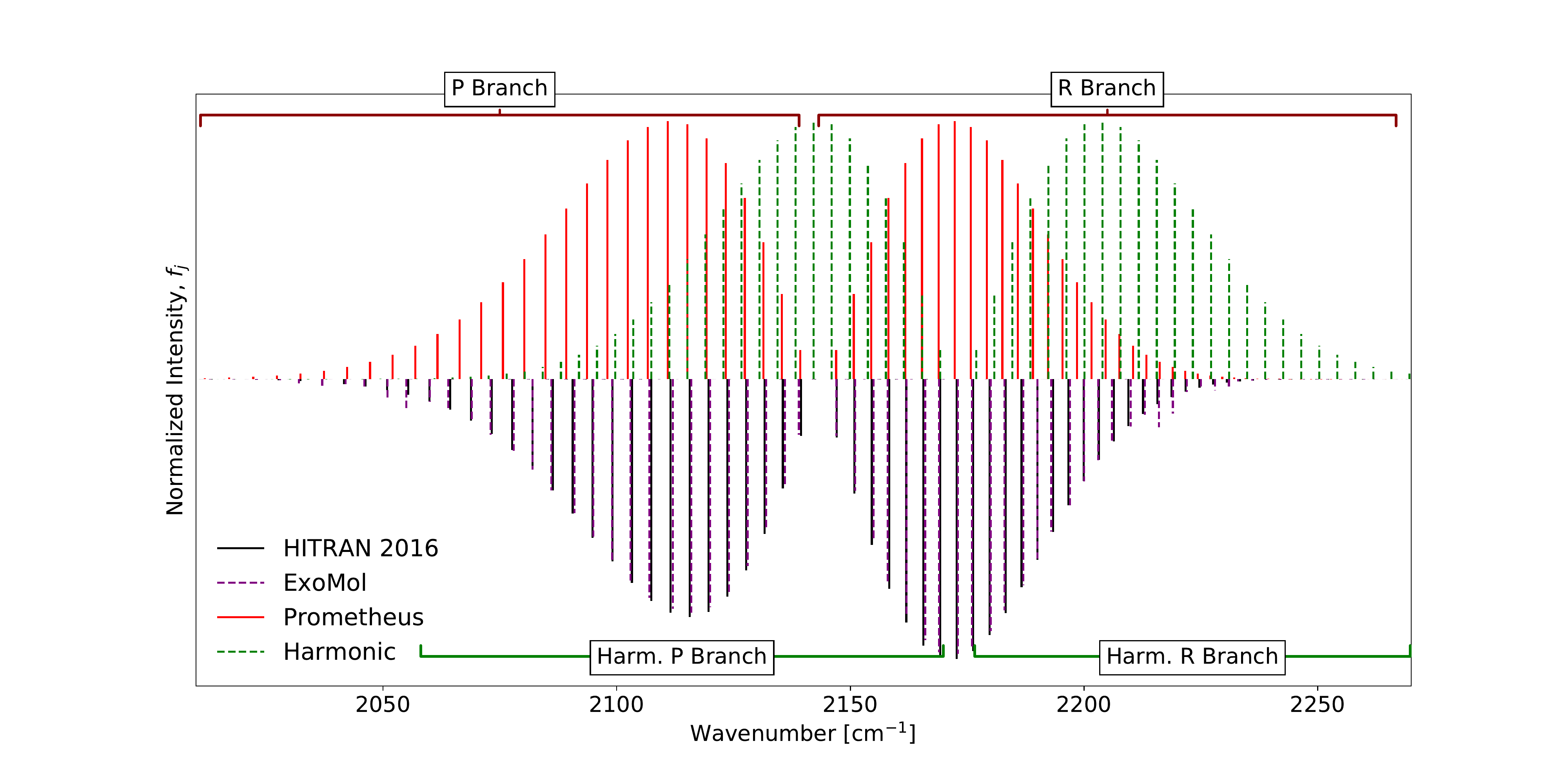}
    \caption{A comparison of the CO spectra produced by Prometheus, the harmonic method, ExoMol \citep[][]{15LiGoRo} and HITRAN 2016 data \citep[][]{17GoRoHi, 04CoHaXX, 18DeBeSu, 15LiGoRo}. The intensities for all sets of data have been normalised and therefore are relative values.}
    \label{fig:CO/CO_Rovibe_Spectrum_Fundamental_PROvHARMvHITRAN2016vExoMol}
\end{figure}

The spectrum produced via the harmonic method (Figure \ref{fig:CO/CO_Rovibe_Spectrum_Fundamental_PROvHARMvHITRAN2016vExoMol}) gives a band origin of $2173.07~\rm{cm}^{-1}$, which has a difference of $29.80~\rm{cm}^{-1}$ to \citet{04CoHaXX}. It is also displaced from ExoMol/HITRAN by approximately 30-40 wavenumbers. Such a displacement naturally renders the harmonic CO spectrum considerably inaccurate when considering the positions of individual $J$ transitions.

In general the distributions of the intensities between HITRAN and Prometheus are similar. On closer inspection, only the fundamental transitions at lower rotational energy levels, $J$, are modelled with high accuracy.

Indeed, most of the lower target transitions lie within a couple of wave numbers of Prometheus’ transitions, but this starts to falter for the higher values of $J$. The error for the higher transitions is then exaggerated due to the slight position differences of the band origin.

The relative intensities appear in a good agreement, although it is worth mentioning that ExoMol is exhibiting some increased intensities for certain higher level transitions. This is not shown by either HITRAN2016 nor Prometheus data.

\subsection{Comparisons of Constants}
As shown in Table \ref{table:comparing_spec_values}, the harmonic approximation consistently describes the ground level rotational constant ($B_0$) well but poorly estimates the upper rotational constant ($B_1$). This makes realistically modelling of any Q branch impossible since it relies on a difference between the ground and excited state rotational constant (see Eq. \ref{eq:28} and Sec.\ref{subsection:Rovibrational_Spectra}). 

The TOSH corrected rotational constants, rather than offering the predicted ``vibrationally averaged" value between to the first and ground states, actually model the first excited state well (see Table \ref{table:comparing_spec_values}). Unfortunately, the TOSH rotational constant overestimates the ground state value, with the harmonic approach typically offering a better approximation. Other studies have also shown \citep[e.g.\ Table 1 in ][]{67EnXXXX} that the uncorrected harmonic approximation is a good approximation for the ground state, whereas for the higher levels this is not the case.

Both the ground and first excited level rotational results have been included in Table \ref{table:comparing_spec_values} and discussed within the previous results sections. This has been done to highlight that using the current modified theory, where we amalgamate the TOSH and harmonic results, is the best option to match reference spectra. The results used for Prometheus are highlighted in the table for each case. 

Typically Prometheus is able to replicate the band origin within a single wavenumber using TOSH, with the exception of H$_2$. Prometheus has also shown it is capable of estimating the band origins to a greater accuracy than that of the harmonic method, often with order of magnitudes improvements.  

Whilst TOSH can approximate the upper rotational constant better than the harmonic model approximates the lower one, both still only approximate the literature values.

This has the further effect of causing incorrect spacing of transitions in the branches, which has been shown for the high-$J$ rotational transitions in all spectra produced. We can verify this conjecture by reviewing the equations \ref{eq:26}-\ref{eq:30}. Each of these position equations have a quadratic dependence on the quantum number $J$, therefore any variation between the rotational constants (such as the differences between the literature and Prometheus), will become more apparent at the higher transitions.

It is also worth noting that we have not used any distortion constants or additional rotation constants in our calculations. We did this to keep the method as simple as possible. The effects of this exclusion is most notable for H$_2$.

\section{Conclusions}
\label{section:Conclusions}
As shown and discussed in section \ref{section:Results}, the TOSH method produces full fundamental rovibrational spectra at higher accuracy than the harmonic approximation for a slight increase in computational complexity. The novelty here is the suitable modeling of the ro-vibrational transition, rather than a simple vibrational modeling of the fundamental. This has never been attempted within the TOSH framework before. 

Provided the anharmonicity constants are of high quality (essentially meaning the potential energy curve they are obtained from is of spectroscopic quality), Prometheus can seemingly replicate a spectrum that is comparable to HITRAN at the lower transitions. Despite the breakdown in accuracy at higher transitions, Prometheus still produces a satisfactory approximation at extremely low computational cost (essentially only the cost of determining the necessary quartic constants: $\eta_{ii}$, $\eta_{iii}$ and $\eta_{iiii}$).

One major caveat is that the current theory for the $\sigma$ shift does not accommodate for two differing vibrational constants for the upper and lower states. Instead the lower value must be calculated via a harmonic methodology, which in turn leads to similar inaccuracies in the transition positions. 

A second issue is the Prometheus spectra do not currently model specific phenomena such as alteration in line intensity due to nuclear spin and the splitting of rotational levels. This arguably is not a crucial element, as the code is currently focussing on line position rather than intensities. In the case of effects due to nuclear spin, the intensities variation is more of a cosmetic issue. The absent lines due to zero spin is arguably a more important effect, and as such has been modelled by Prometheus.

A recurring note for improvement throughout this paper has pertained to the shift variable, $\sigma$. It is a useful tool to use to account for anharmonic corrections of the rotational constant, however in its current form it is only suitable for the fundamental transition. To improve the spectra, we need to modify the theory to obtain a shift that varies with vibrational level instead of being optimised for the fundamental, $0 \rightarrow 1$, vibrational transition.

Another aspect of further work would revolve around creating a more robust means of evaluating the additional effects arising from symmetry of a molecule. This would include considering the phenomena arising due to properties such a spin and triplet structure. 

Finally, the next stage is to adapt the theory to triatomic molecules. Once triatomics have been successfully modelled we can then progress to modelling larger polyatomic molecules.

\section*{Acknowledgements}

We would like to thank our colleagues in the Milne Centre for all their input and contributions. We also thank Dr.~M.~Gorman and Miss V.~H.~J.~Clark for their feedback and input on the manuscript, in addition to the most helpful guidance provided by the referee.
We acknowledge the support of JINA-CEE (NSF Grant PHY-1430152) and STFC (through the University of Hull’s Consolidated Grant ST/R000840/1), and ongoing access to {\tt viper}, the University of Hull High Performance Computing Facility. MP acknowledges support from the ``Lend{\"u}let-2014" Programme of the Hungarian Academy of Sciences (Hungary), and the ERC Consolidator Grant (Hungary) programme (RADIOSTAR, G.A. n. 724560). This work was supported by the European Union’s Horizon 2020 research and innovation programme (ChETEC-INFRA -- Project no. 101008324), and the IReNA network supported by US NSF AccelNet.

%


\bibliographystyle{ApJ}
\bibliography{Main_body}

\begin{thebibliography}{}
\expandafter\ifx\csname natexlab\endcsname\relax\def\natexlab#1{#1}\fi

\bibitem[{Agundez \& Cernicharo(2006)}]{06AGSeXX}
Agundez, M., \& Cernicharo, J. 2006, \apj, 650, 374

\bibitem[{{Aladro} {et~al.}(2015){Aladro}, {Mart{\'\i}n}, {Riquelme}, {Henkel},
  {Mauersberger}, {Mart{\'\i}n-Pintado}, {Wei{\ss}}, {Lefevre}, {Kramer},
  {Requena-Torres}, \& {Armijos-Abenda{\~n}o}}]{aladro:15}
{Aladro}, R., {Mart{\'\i}n}, S., {Riquelme}, D., {et~al.} 2015, \aap, 579, A101

\bibitem[{{Bacmann} {et~al.}(2012){Bacmann}, {Taquet}, {Faure}, {Kahane}, \&
  {Ceccarelli}}]{bacmann:12}
{Bacmann}, A., {Taquet}, V., {Faure}, A., {Kahane}, C., \& {Ceccarelli}, C.
  2012, \aap, 541, L12

\bibitem[{Bains {et~al.}(2020)Bains, Petkowski, Seager, Ranjan, Sousa-Silva,
  Rimmer, Zhan, Greaves, \& Richards}]{20BaPeSe}
Bains, W., Petkowski, J.~J., Seager, S., {et~al.} 2020, arXiv, arXiv:2009.06499

\bibitem[{Barstow {et~al.}(2015)Barstow, Aigrain, Irwin, Kendrew, \&
  Fletcher}]{15BaAiIr}
Barstow, J.~K., Aigrain, S., Irwin, P. G.~J., Kendrew, S., \& Fletcher, L.~N.
  2015, \mnras, 448, 2546

\bibitem[{Be{\'{c}} {et~al.}(2016)Be{\'{c}}, Futami, W{\'{o}}jcik, \&
  Ozaki}]{16BeFuWo}
Be{\'{c}}, K., Futami, Y., W{\'{o}}jcik, M., \& Ozaki, Y. 2016, Phys. Chem.
  Chem. Phys., 18, doi:10.1039/C6CP00924G

\bibitem[{{Bendtsen} \& {Rasmussen}(2000)}]{00BeRaXX}
{Bendtsen}, J., \& {Rasmussen}, F. 2000, J. Raman Spectrosc., 31, 433

\bibitem[{Brown {et~al.}(2013)Brown, Sung, Benner, Devi, Boudon, Gabard,
  Wenger, Campargue, Leshchishina, Kassi, Mondelain, Wang, Daumont,
  R{\'{e}}galia, Rey, Thomas, Tyuterev, Lyulin, Nikitin, Niederer, Albert,
  Bauerecker, Quack, O'Brien, Gordon, Rothman, Sasada, Coustenis, Smith,
  Carrington, Wang, Mantz, \& Spickler}]{13BrSuBe}
Brown, L., Sung, K., Benner, D., {et~al.} 2013, \jqsrt, 130, 201

\bibitem[{Campargue {et~al.}(2012)Campargue, Kassi, Pachucki, \&
  Komasa}]{12CaKaPa}
Campargue, A., Kassi, S., Pachucki, K., \& Komasa, J. 2012, Phys. Chem. Chem.
  Phys., 14, 802

\bibitem[{Chubb {et~al.}(2021)Chubb, Rocchetto, Yurchenko, Min, Waldmann,
  Barstow, Molli{\`{e}}re, Al-Refaie, Phillips, \& Tennyson}]{21ChRoYu}
Chubb, K.~L., Rocchetto, M., Yurchenko, S.~N., {et~al.} 2021, A\&A, 646, A21

\bibitem[{Clark \& Benoit(2020)}]{19ClBeXX}
Clark, V.~H., \& Benoit, D.~M. 2020, Proc Int Astron Union, 468

\bibitem[{Coles {et~al.}(2019)Coles, Yurchenko, \& Tennyson}]{19CoYuTe}
Coles, P.~A., Yurchenko, S.~N., \& Tennyson, J. 2019, \mnras, 490, 4638

\bibitem[{{Col{\'o}n} {et~al.}(2020){Col{\'o}n}, {Kreidberg}, {Welbanks},
  {Line}, {Madhusudhan}, {Beatty}, {Tamburo}, {Stevenson}, {Mandell},
  {Rodriguez}, {Barclay}, {Lopez}, {Stassun}, {Angerhausen}, {Fortney},
  {James}, {Pepper}, {Ahlers}, {Plavchan}, {Awiphan}, {Kotnik}, {McLeod},
  {Murawski}, {Chotani}, {LeBrun}, {Matzko}, {Rea}, {Vidaurri}, {Webster},
  {Williams}, {Cox}, {Tan}, \& {Gilbert}}]{2020AJ....160..280C}
{Col{\'o}n}, K.~D., {Kreidberg}, L., {Welbanks}, L., {et~al.} 2020, \aj, 160,
  280

\bibitem[{{Costagliola} {et~al.}(2011){Costagliola}, {Aalto}, {Rodriguez},
  {Muller}, {Spoon}, {Mart{\'\i}n}, {Per{\'e}z-Torres}, {Alberdi}, {Lindberg},
  {Batejat}, {J{\"u}tte}, {van der Werf}, \& {Lahuis}}]{costagliola:11}
{Costagliola}, F., {Aalto}, S., {Rodriguez}, M.~I., {et~al.} 2011, \aap, 528,
  A30

\bibitem[{Cox(2002)}]{02CoXXXX}
Cox, A.~N., ed. 2002, {Allen's Astrophysical Quantities} (New York, NY:
  Springer New York), doi:10.1007/978-1-4612-1186-0

\bibitem[{{Coxon} \& {Hajigeorgiou}(2004)}]{04CoHaXX}
{Coxon}, J.~A., \& {Hajigeorgiou}, P.~G. 2004, \jcp, 121, 2992

\bibitem[{{Cristallo} {et~al.}(2015){Cristallo}, {Straniero}, {Piersanti}, \&
  {Gobrecht}}]{cristallo:15}
{Cristallo}, S., {Straniero}, O., {Piersanti}, L., \& {Gobrecht}, D. 2015,
  \apjs, 219, 40

\bibitem[{Damiano {et~al.}(2017)Damiano, Morello, Tsiaras, Zingales, \&
  Tinetti}]{17DaMoTs}
Damiano, M., Morello, G., Tsiaras, A., Zingales, T., \& Tinetti, G. 2017, \apj,
  154, 39

\bibitem[{de~Kok {et~al.}(2013)de~Kok, Brogi, Snellen, Birkby, Albrecht, \&
  de~Mooij}]{13KoBrSn}
de~Kok, R., Brogi, M., Snellen, I., {et~al.} 2013, A\&A, 554, A82

\bibitem[{{Des Marais} {et~al.}(2008){Des Marais}, {Nuth}, {Allamandola},
  {Boss}, {Farmer}, {Hoehler}, {Jakosky}, {Meadows}, {Pohorille}, {Runnegar},
  \& {Spormann}}]{2008AsBio...8..715D}
{Des Marais}, D.~J., {Nuth}, Joseph~A., I., {Allamandola}, L.~J., {et~al.}
  2008, Astrobiology, 8, 715

\bibitem[{{Devi} {et~al.}(2018){Devi}, {Benner}, {Sung}, {Crawford}, {Li},
  {Gamache}, {Smith}, {Gordon}, \& {Mantz}}]{18DeBeSu}
{Devi}, V.~M., {Benner}, D.~C., {Sung}, K., {et~al.} 2018, \jqsrt, 218, 203

\bibitem[{Domagal-Goldman {et~al.}(2014)Domagal-Goldman, Segura, Claire,
  Robinson, \& Meadows}]{14DoSeCl}
Domagal-Goldman, S.~D., Segura, A., Claire, M.~W., Robinson, T.~D., \& Meadows,
  V.~S. 2014, \apj, 792, doi:10.1088/0004-637X/792/2/90

\bibitem[{{Edwards} {et~al.}(1993){Edwards}, {Roncin}, {Launay}, \&
  {Rostas}}]{93EdRoLa}
{Edwards}, S., {Roncin}, J.~Y., {Launay}, F., \& {Rostas}, F. 1993, J. Mol.
  Spectrosc., 162, 257

\bibitem[{{Endres} {et~al.}(2016){Endres}, {Schlemmer}, {Schilke}, {Stutzki},
  \& {M{\"u}ller}}]{2016JMoSp.327...95E}
{Endres}, C.~P., {Schlemmer}, S., {Schilke}, P., {Stutzki}, J., \&
  {M{\"u}ller}, H. S.~P. 2016, J. Mol. Spectrosc., 327, 95

\bibitem[{Endres(1967)}]{67EnXXXX}
Endres, P.~F. 1967, J. Chem. Phys., 47, 798

\bibitem[{{Evans} {et~al.}(2018){Evans}, {Sing}, {Goyal}, {Nikolov}, {Marley},
  {Zahnle}, {Henry}, {Barstow}, {Alam}, {Sanz-Forcada}, {Kataria}, {Lewis},
  {Lavvas}, {Ballester}, {Ben-Jaffel}, {Blumenthal}, {Bourrier}, {Drummond},
  {Garc{\'\i}a Mu{\~n}oz}, {L{\'o}pez-Morales}, {Tremblin}, {Ehrenreich},
  {Wakeford}, {Buchhave}, {Lecavelier des Etangs}, {H{\'e}brard}, \&
  {Williamson}}]{2018AJ....156..283E}
{Evans}, T.~M., {Sing}, D.~K., {Goyal}, J.~M., {et~al.} 2018, \aj, 156, 283

\bibitem[{Falk \& Whalley(1961)}]{61FaWhXX}
Falk, M., \& Whalley, E. 1961, J. Chem. Phys., 34, 1554

\bibitem[{Fortenberry \& Lee(2019)}]{19FoLeXX}
Fortenberry, R.~C., \& Lee, T.~J. 2019, {Computational vibrational spectroscopy
  for the detection of molecules in space}, 1st edn., Vol.~15 (Elsevier B.V.),
  173--202, doi:10.1016/bs.arcc.2019.08.006

\bibitem[{Franz {et~al.}(2017)Franz, Trainer, Malespin, Mahaffy, Atreya,
  Becker, Benna, Conrad, Eigenbrode, Freissinet, Manning, Prats, Raaen, \&
  Wong}]{17FrTrMa}
Franz, H.~B., Trainer, M.~G., Malespin, C.~A., {et~al.} 2017, P\&SS, 138, 44

\bibitem[{{Galli} \& {Palla}(1998)}]{galli:98}
{Galli}, D., \& {Palla}, F. 1998, \aap, 335, 403

\bibitem[{Gamache {et~al.}(1998)Gamache, Goldman, \& Rothman}]{98GaGoRo}
Gamache, R., Goldman, A., \& Rothman, L. 1998, \jqsrt, 59, 495

\bibitem[{{Gordon} {et~al.}(2017){Gordon}, {Rothman}, {Hill}, {Kochanov},
  {Tan}, {Bernath}, {Birk}, {Boudon}, {Campargue}, {Chance}, {Drouin}, {Flaud},
  {Gamache}, {Hodges}, {Jacquemart}, {Perevalov}, {Perrin}, {Shine}, {Smith},
  {Tennyson}, {Toon}, {Tran}, {Tyuterev}, {Barbe}, {Cs{\'a}sz{\'a}r}, {Devi},
  {Furtenbacher}, {Harrison}, {Hartmann}, {Jolly}, {Johnson}, {Karman},
  {Kleiner}, {Kyuberis}, {Loos}, {Lyulin}, {Massie}, {Mikhailenko},
  {Moazzen-Ahmadi}, {M{\"u}ller}, {Naumenko}, {Nikitin}, {Polyansky}, {Rey},
  {Rotger}, {Sharpe}, {Sung}, {Starikova}, {Tashkun}, {Auwera}, {Wagner},
  {Wilzewski}, {Wcis{\l}o}, {Yu}, \& {Zak}}]{17GoRoHi}
{Gordon}, I.~E., {Rothman}, L.~S., {Hill}, C., {et~al.} 2017, \jqsrt, 203, 3

\bibitem[{{Greaves} {et~al.}(2020){Greaves}, {Richards}, {Bains}, {Rimmer},
  {Sagawa}, {Clements}, {Seager}, {Petkowski}, {Sousa-Silva}, {Ranjan},
  {Drabek-Maunder}, {Fraser}, {Cartwright}, {Mueller-Wodarg}, {Zhan},
  {Friberg}, {Coulson}, {Lee}, \& {Hoge}}]{greaves:20}
{Greaves}, J.~S., {Richards}, A. M.~S., {Bains}, W., {et~al.} 2020, Nat.
  Astron., arXiv:2009.06593

\bibitem[{Greaves {et~al.}(2020)Greaves, Richards, Bains, Rimmer, Sagawa,
  Clements, Seager, Petkowski, Sousa-Silva, Ranjan, Drabek-Maunder, Fraser,
  Cartwright, Mueller-Wodarg, Zhan, Friberg, Coulson, Lee, \& Hoge}]{20GrRiBa}
Greaves, J.~S., Richards, A.~M., Bains, W., {et~al.} 2020, Nat. Astron.,
  arXiv:2009.06593

\bibitem[{Guilluy {et~al.}(2019)Guilluy, Sozzetti, Brogi, Bonomo, Giacobbe,
  Claudi, \& Benatti}]{19GuSoBr}
Guilluy, G., Sozzetti, A., Brogi, M., {et~al.} 2019, A\&A, 625, A107

\bibitem[{Habart {et~al.}(2005)Habart, Walmsley, Verstraete, Cazaux, Maiolino,
  Cox, Boulanger, \& {Des For{\^{e}}ts}}]{04HaWaVe}
Habart, E., Walmsley, M., Verstraete, L., {et~al.} 2005, Space Sci. Rev., 119,
  71

\bibitem[{Hanson-Heine(2019)}]{19HaXXXX}
Hanson-Heine, M. W.~D. 2019, J. Phys. Chem. A, 123, 9800

\bibitem[{Harada {et~al.}(2018)Harada, Sakamoto, Mart{\'{i}}n, Aalto, Aladro,
  \& Sliwa}]{18HaSaMa}
Harada, N., Sakamoto, K., Mart{\'{i}}n, S., {et~al.} 2018, \apj, 855, 49

\bibitem[{{Hedrosa} {et~al.}(2013){Hedrosa}, {Abia}, {Busso}, {Cristallo},
  {Dom{\'\i}nguez}, {Palmerini}, {Plez}, \& {Straniero}}]{hedrosa:13}
{Hedrosa}, R.~P., {Abia}, C., {Busso}, M., {et~al.} 2013, \apjl, 768, L11

\bibitem[{Herzberg(1949)}]{49XXXX}
Herzberg, G. 1949, Nature, 163, 170

\bibitem[{{Horner}(2000)}]{00HoXXXX}
{Horner}, J.~M. 2000, Rev. Environ. Health, 15(3), 289–298

\bibitem[{{Jennings} \& {Brault}(1983)}]{83JeBrXX}
{Jennings}, D.~E., \& {Brault}, J.~W. 1983, J. Mol. Spectrosc., 102, 265

\bibitem[{Jo {et~al.}(2017)Jo, Seon, Min, Edelstein, \& Han}]{17JoSeMi}
Jo, Y.-S., Seon, K.-I., Min, K.-W., Edelstein, J., \& Han, W. 2017, Astrophys.
  J., Suppl. Ser., 231, 21

\bibitem[{{Karakas} \& {Lattanzio}(2014)}]{karakas:14}
{Karakas}, A.~I., \& {Lattanzio}, J.~C. 2014, \pasa, 31, e030

\bibitem[{Katyal {et~al.}(2019)Katyal, Nikolaou, Godolt, Grenfell, Tosi,
  Schreier, \& Rauer}]{19KaNiGo}
Katyal, N., Nikolaou, A., Godolt, M., {et~al.} 2019, \apj, 875, 31

\bibitem[{Komasa {et~al.}(2011)Komasa, Piszczatowski, Lach, Przybytek,
  Jeziorski, \& Pachucki}]{11KoPiLa}
Komasa, J., Piszczatowski, K., Lach, G., {et~al.} 2011, J. Chem. Theory
  Comput., 7, 3105

\bibitem[{{Lammer} {et~al.}(2019){Lammer}, {Spro{\ss}}, {Grenfell}, {Scherf},
  {Fossati}, {Lendl}, \& {Cubillos}}]{19LaSpGr}
{Lammer}, H., {Spro{\ss}}, L., {Grenfell}, J.~L., {et~al.} 2019, Astrobiology,
  19, 927

\bibitem[{Li {et~al.}(2015)Li, Gordon, Rothman, Tan, Hu, Kassi, Campargue, \&
  Medvedev}]{15LiGoRo}
Li, G., Gordon, I.~E., Rothman, L.~S., {et~al.} 2015, ApJS, 216, 15

\bibitem[{Li \& Roy(2007)}]{07LiLeXX}
Li, H., \& Roy, R.~L. 2007, J. Chem. Phys, 126, 224301

\bibitem[{{Li} {et~al.}(2013){Li}, {Heays}, {Visser}, {Ubachs}, {Lewis},
  {Gibson}, \& {van Dishoeck}}]{13LiHeVi}
{Li}, X., {Heays}, A.~N., {Visser}, R., {et~al.} 2013, \aap, 555, A14

\bibitem[{Lin {et~al.}(2008)Lin, Gilbert, \& Gill}]{08LiGiGI}
Lin, C.~Y., Gilbert, A. T.~B., \& Gill, P. M.~W. 2008, Theor. Chem. Acc., 120,
  23

\bibitem[{Luspay-Kuti {et~al.}(2018)Luspay-Kuti, Mousis, Lunine, Ellinger,
  Pauzat, Raut, Bouquet, Mandt, Maggiolo, Ronnet, Brugger, Ozgurel, \&
  Fuselier}]{18LuMoLu}
Luspay-Kuti, A., Mousis, O., Lunine, J.~I., {et~al.} 2018, Space Sci. Rev.,
  214, 1

\bibitem[{{Malathy Devi} {et~al.}(2014){Malathy Devi}, Kleiner, Sams, Brown,
  Benner, \& Fletcher}]{14DeKlSa}
{Malathy Devi}, V., Kleiner, I., Sams, R.~L., {et~al.} 2014, J. Mol.
  Spectrosc., 298, 11

\bibitem[{McKemmish {et~al.}(2019)McKemmish, Masseron, Hoeijmakers,
  P{\'{e}}rez-Mesa, Grimm, Yurchenko, \& Tennyson}]{19McMaHo}
McKemmish, L.~K., Masseron, T., Hoeijmakers, H.~J., {et~al.} 2019, \mnras, 488,
  2836

\bibitem[{{Meadows}(2017)}]{17MeXXXX}
{Meadows}, V.~S. 2017, Astrobiology, 17, 1022

\bibitem[{Meadows {et~al.}(2018)Meadows, Reinhard, Arney, Parenteau,
  Schwieterman, Domagal-Goldman, Lincowski, Stapelfeldt, Rauer, DasSarma,
  Hegde, Narita, Deitrick, Lustig-Yaeger, Lyons, Siegler, \&
  Grenfell}]{18MeReAr}
Meadows, V.~S., Reinhard, C.~T., Arney, G.~N., {et~al.} 2018, Astrobiology, 18,
  630

\bibitem[{Meshkov {et~al.}(2018)Meshkov, Stolyarov, Ermilov, Medvedev, Ushakov,
  \& Gordon}]{18MeStEr}
Meshkov, V.~V., Stolyarov, A.~V., Ermilov, A.~Y., {et~al.} 2018, \jqsrt, 217,
  262

\bibitem[{Morales {et~al.}(1998)Morales, Espinosa, Borrell, Tapanes, Estrada,
  \& {De la Fuente}}]{08GaWeHe}
Morales, R., Espinosa, G., Borrell, Y.~J., {et~al.} 1998, Biotecnol. Apl, 15,
  149

\bibitem[{{Muller} {et~al.}(2006){Muller}, {Gu{\'e}lin}, {Dumke}, {Lucas}, \&
  {Combes}}]{muller:06}
{Muller}, S., {Gu{\'e}lin}, M., {Dumke}, M., {Lucas}, R., \& {Combes}, F. 2006,
  \aap, 458, 417

\bibitem[{{Nomoto} {et~al.}(2013){Nomoto}, {Kobayashi}, \&
  {Tominaga}}]{nomoto:13}
{Nomoto}, K., {Kobayashi}, C., \& {Tominaga}, N. 2013, \araa, 51, 457

\bibitem[{Oyama {et~al.}(1980)Oyama, Carle, Woeller, Pollack, Reynolds, \&
  Craig}]{80OyCaWo}
Oyama, V.~I., Carle, G.~C., Woeller, F., {et~al.} 1980, J. Geophys. Res., 85,
  7891

\bibitem[{Paw{\l}owski {et~al.}(2002)Paw{\l}owski, J{\o}rgensen, Olsen,
  Hegelund, Helgaker, Gauss, Bak, \& Stanton}]{01PaJoOl}
Paw{\l}owski, F., J{\o}rgensen, P., Olsen, J., {et~al.} 2002, J. Chem. Phys.,
  116, 6482

\bibitem[{{Pearson} {et~al.}(2005){Pearson}, {Drouin}, \&
  {Pickett}}]{2005IAUS..231P.270P}
{Pearson}, J.~C., {Drouin}, B.~J., \& {Pickett}, H.~M. 2005, in Astrochemistry:
  Recent Successes and Current Challenges, ed. D.~C. {Lis}, G.~A. {Blake}, \&
  E.~{Herbst}, Vol. 231, 270

\bibitem[{{Pignatari} {et~al.}(2016){Pignatari}, {Herwig}, {Hirschi},
  {Bennett}, {Rockefeller}, {Fryer}, {Timmes}, {Ritter}, {Heger}, {Jones},
  {Battino}, {Dotter}, {Trappitsch}, {Diehl}, {Frischknecht}, {Hungerford},
  {Magkotsios}, {Travaglio}, \& {Young}}]{pignatari:16}
{Pignatari}, M., {Herwig}, F., {Hirschi}, R., {et~al.} 2016, \apjs, 225, 24

\bibitem[{Polyansky {et~al.}(2018)Polyansky, Kyuberis, Zobov, Tennyson,
  Yurchenko, \& Lodi}]{18PoKyZo}
Polyansky, O.~L., Kyuberis, A.~A., Zobov, N.~F., {et~al.} 2018, \mnras, 480,
  2597

\bibitem[{{Rauscher} {et~al.}(2002){Rauscher}, {Heger}, {Hoffman}, \&
  {Woosley}}]{rauscher:02}
{Rauscher}, T., {Heger}, A., {Hoffman}, R.~D., \& {Woosley}, S.~E. 2002, \apj,
  576, 323

\bibitem[{Rees(1947)}]{47ReXXXX}
Rees, A.~L. 1947, Proc. Phys. Soc., 59, 998

\bibitem[{{Ritter} {et~al.}(2018){Ritter}, {Herwig}, {Jones}, {Pignatari},
  {Fryer}, \& {Hirschi}}]{ritter:18}
{Ritter}, C., {Herwig}, F., {Jones}, S., {et~al.} 2018, \mnras, 480, 538

\bibitem[{{Rivilla} {et~al.}(2016){Rivilla}, {Fontani}, {Beltr{\'a}n},
  {Vasyunin}, {Caselli}, {Mart{\'\i}n-Pintado}, \& {Cesaroni}}]{rivilla:16}
{Rivilla}, V.~M., {Fontani}, F., {Beltr{\'a}n}, M.~T., {et~al.} 2016, \apj,
  826, 161

\bibitem[{Roueff {et~al.}(2019)Roueff, Abgrall, Czachorowski, Pachucki,
  Puchalski, \& Komasa}]{19RoAbCz}
Roueff, E., Abgrall, H., Czachorowski, P., {et~al.} 2019, A\&A, 630, A58

\bibitem[{Roy {et~al.}(2006)Roy, Huang, \& Jary}]{06LeHuJa}
Roy, R.~L., Huang, Y., \& Jary, C. 2006, J. Chem. Phys, 125, 164310

\bibitem[{Schlegel(1974)}]{73ScXXXX}
Schlegel, H.~G. 1974, Tellus, 26, 11

\bibitem[{Schwartz \& {Le Roy}(1987)}]{87ScLeXX}
Schwartz, C., \& {Le Roy}, R.~J. 1987, J. Mol. Spectrosc., 121, 420

\bibitem[{{Seager} {et~al.}(2016){Seager}, {Bains}, \& {Petkowski}}]{16SeBaPe}
{Seager}, S., {Bains}, W., \& {Petkowski}, J.~J. 2016, Astrobiology, 16, 465

\bibitem[{Seager {et~al.}(2020)Seager, Huang, Petkowski, \&
  Pajusalu}]{20SeHuPe}
Seager, S., Huang, J., Petkowski, J.~J., \& Pajusalu, M. 2020, Nat. Astron, 4,
  802

\bibitem[{{Shimajiri} {et~al.}(2017){Shimajiri}, {Andr{\'e}}, {Braine},
  {K{\"o}nyves}, {Schneider}, {Bontemps}, {Ladjelate}, {Roy}, {Gao}, \&
  {Chen}}]{shimajiri:17}
{Shimajiri}, Y., {Andr{\'e}}, P., {Braine}, J., {et~al.} 2017, \aap, 604, A74

\bibitem[{Sousa-Silva {et~al.}(2015)Sousa-Silva, Al-Refaie, Tennyson, \&
  Yurchenko}]{14SoAlTe}
Sousa-Silva, C., Al-Refaie, A.~F., Tennyson, J., \& Yurchenko, S.~N. 2015,
  \mnras, 446, 2337

\bibitem[{Sousa-Silva {et~al.}(2019)Sousa-Silva, Petkowski, \&
  Seager}]{19SoPeSe}
Sousa-Silva, C., Petkowski, J.~J., \& Seager, S. 2019, Phys. Chem. Chem. Phys.,
  21, 18970

\bibitem[{Sousa-Silva {et~al.}(2013)Sousa-Silva, Yurchenko, \&
  Tennyson}]{13SoYuTe}
Sousa-Silva, C., Yurchenko, S.~N., \& Tennyson, J. 2013, J. Mol. Spectrosc.,
  288, 28

\bibitem[{Spro{\ss} {et~al.}(2018)Spro{\ss}, Lammer, {Lee Grenfell}, Scherf,
  Fossati, Lendl, \& Cubillos}]{18SpLaGr}
Spro{\ss}, L., Lammer, H., {Lee Grenfell}, J., {et~al.} 2018, EPSC, 12, 2018

\bibitem[{Strobel \& Shemansky(1982)}]{82StShXX}
Strobel, D.~F., \& Shemansky, D.~E. 1982, J. Geophys. Res. Space Phys., 87,
  1361

\bibitem[{{Sukhbold} {et~al.}(2016){Sukhbold}, {Ertl}, {Woosley}, {Brown}, \&
  {Janka}}]{sukhbold:16}
{Sukhbold}, T., {Ertl}, T., {Woosley}, S.~E., {Brown}, J.~M., \& {Janka}, H.~T.
  2016, \apj, 821, 38

\bibitem[{{Swain} {et~al.}(2009){Swain}, {Tinetti}, {Vasisht}, {Deroo},
  {Griffith}, {Bouwman}, {Chen}, {Yung}, {Burrows}, {Brown}, {Matthews},
  {Rowe}, {Kuschnig}, \& {Angerhausen}}]{swain:09}
{Swain}, M.~R., {Tinetti}, G., {Vasisht}, G., {et~al.} 2009, \apj, 704, 1616

\bibitem[{Tennyson {et~al.}(2016)Tennyson, Yurchenko, Al-Refaie, Barton, Chubb,
  Coles, Diamantopoulou, Gorman, Hill, Lam, Lodi, McKemmish, Na, Owens,
  Polyansky, Rivlin, Sousa-Silva, Underwood, Yachmenev, \& Zak}]{16TeYuAl}
Tennyson, J., Yurchenko, S.~N., Al-Refaie, A.~F., {et~al.} 2016, J. Mol.
  Spectrosc., 327, 73

\bibitem[{Tennyson {et~al.}(2020)Tennyson, Yurchenko, Al-Refaie, Clark, Chubb,
  Conway, Dewan, Gorman, Hill, Lynas-Gray, Mellor, McKemmish, Owens, Polyansky,
  Semenov, Somogyi, Tinetti, Upadhyay, Waldmann, Wang, Wright, \&
  Yurchenko}]{20TeYuAl}
---. 2020, \jqsrt, 255, 107228

\bibitem[{Tessenyi {et~al.}(2013)Tessenyi, Tinetti, Savini, \&
  Pascale}]{13TeTiSa}
Tessenyi, M., Tinetti, G., Savini, G., \& Pascale, E. 2013, \icarus, 226, 1654

\bibitem[{{Tinetti} {et~al.}(2013){Tinetti}, {Encrenaz}, \&
  {Coustenis}}]{tinetti:13}
{Tinetti}, G., {Encrenaz}, T., \& {Coustenis}, A. 2013, \aapr, 21, 63

\bibitem[{{Tinetti} {et~al.}(2018){Tinetti}, {Drossart}, {Eccleston},
  {Hartogh}, {Heske}, {Leconte}, {Micela}, {Ollivier}, {Pilbratt}, {Puig},
  {Turrini}, {Vandenbussche}, {Wolkenberg}, {Beaulieu}, {Buchave}, {Ferus},
  {Griffin}, {Guedel}, {Justtanont}, {Lagage}, {Machado}, {Malaguti}, {Min},
  {N{\o}rgaard-Nielsen}, {Rataj}, {Ray}, {Ribas}, {Swain}, {Szabo}, {Werner},
  {Barstow}, {Burleigh}, {Cho}, {du Foresto}, {Coustenis}, {Decin}, {Encrenaz},
  {Galand}, {Gillon}, {Helled}, {Morales}, {Mu{\~n}oz}, {Moneti}, {Pagano},
  {Pascale}, {Piccioni}, {Pinfield}, {Sarkar}, {Selsis}, {Tennyson}, {Triaud},
  {Venot}, {Waldmann}, {Waltham}, {Wright}, {Amiaux}, {Augu{\`e}res},
  {Berth{\'e}}, {Bezawada}, {Bishop}, {Bowles}, {Coffey}, {Colom{\'e}},
  {Crook}, {Crouzet}, {Da Peppo}, {Sanz}, {Focardi}, {Frericks}, {Hunt},
  {Kohley}, {Middleton}, {Morgante}, {Ottensamer}, {Pace}, {Pearson},
  {Stamper}, {Symonds}, {Rengel}, {Renotte}, {Ade}, {Affer}, {Alard}, {Allard},
  {Altieri}, {Andr{\'e}}, {Arena}, {Argyriou}, {Aylward}, {Baccani}, {Bakos},
  {Banaszkiewicz}, {Barlow}, {Batista}, {Bellucci}, {Benatti}, {Bernardi},
  {B{\'e}zard}, {Blecka}, {Bolmont}, {Bonfond}, {Bonito}, {Bonomo}, {Brucato},
  {Brun}, {Bryson}, {Bujwan}, {Casewell}, {Charnay}, {Pestellini}, {Chen},
  {Ciaravella}, {Claudi}, {Cl{\'e}dassou}, {Damasso}, {Damiano}, {Danielski},
  {Deroo}, {Di Giorgio}, {Dominik}, {Doublier}, {Doyle}, {Doyon}, {Drummond},
  {Duong}, {Eales}, {Edwards}, {Farina}, {Flaccomio}, {Fletcher}, {Forget},
  {Fossey}, {Fr{\"a}nz}, {Fujii}, {Garc{\'\i}a-Piquer}, {Gear}, {Geoffray},
  {G{\'e}rard}, {Gesa}, {Gomez}, {Graczyk}, {Griffith}, {Grodent}, {Guarcello},
  {Gustin}, {Hamano}, {Hargrave}, {Hello}, {Heng}, {Herrero}, {Hornstrup},
  {Hubert}, {Ida}, {Ikoma}, {Iro}, {Irwin}, {Jarchow}, {Jaubert}, {Jones},
  {Julien}, {Kameda}, {Kerschbaum}, {Kervella}, {Koskinen}, {Krijger}, {Krupp},
  {Lafarga}, {Landini}, {Lellouch}, {Leto}, {Luntzer}, {Rank-L{\"u}ftinger},
  {Maggio}, {Maldonado}, {Maillard}, {Mall}, {Marquette}, {Mathis}, {Maxted},
  {Matsuo}, {Medvedev}, {Miguel}, {Minier}, {Morello}, {Mura}, {Narita},
  {Nascimbeni}, {Nguyen Tong}, {Noce}, {Oliva}, {Palle}, {Palmer}, {Pancrazzi},
  {Papageorgiou}, {Parmentier}, {Perger}, {Petralia}, {Pezzuto},
  {Pierrehumbert}, {Pillitteri}, {Piotto}, {Pisano}, {Prisinzano}, {Radioti},
  {R{\'e}ess}, {Rezac}, {Rocchetto}, {Rosich}, {Sanna}, {Santerne}, {Savini},
  {Scandariato}, {Sicardy}, {Sierra}, {Sindoni}, {Skup}, {Snellen}, {Sobiecki},
  {Soret}, {Sozzetti}, {Stiepen}, {Strugarek}, {Taylor}, {Taylor}, {Terenzi},
  {Tessenyi}, {Tsiaras}, {Tucker}, {Valencia}, {Vasisht}, {Vazan}, {Vilardell},
  {Vinatier}, {Viti}, {Waters}, {Wawer}, {Wawrzaszek}, {Whitworth}, {Yung},
  {Yurchenko}, {Osorio}, {Zellem}, {Zingales}, \& {Zwart}}]{tinetti:18}
{Tinetti}, G., {Drossart}, P., {Eccleston}, P., {et~al.} 2018, Exp. Astron.,
  46, 135

\bibitem[{Tran {et~al.}(2006)Tran, Flaud, Fouchet, Gabard, \&
  Hartmann}]{06TrFlFo}
Tran, H., Flaud, P.~M., Fouchet, T., Gabard, T., \& Hartmann, J.~M. 2006,
  \jqsrt, 101, 306

\bibitem[{Vangioni {et~al.}(2018)Vangioni, Dvorkin, Olive, Dubois, Molaro,
  Petitjean, Silk, \& Kimm}]{18VaDvOl}
Vangioni, E., Dvorkin, I., Olive, K.~A., {et~al.} 2018, \mnras, 477, 56

\bibitem[{V{\'{a}}zquez \& Stanton(2006)}]{06VaStXX}
V{\'{a}}zquez, J., \& Stanton, J.~F. 2006, Mol Phys, 104, 377

\bibitem[{Viti {et~al.}(1997)Viti, Tennyson, \& Polyansky}]{97ViTePo}
Viti, S., Tennyson, J., \& Polyansky, O.~L. 1997, \mnras, 287, 79

\bibitem[{{Wakelam} \& {Herbst}(2008)}]{wakelam:08}
{Wakelam}, V., \& {Herbst}, E. 2008, \apj, 680, 371

\bibitem[{Wakelam {et~al.}(2010)Wakelam, Smith, Herbst, Troe, Geppert,
  Linnartz, {\"{O}}berg, Roueff, Ag{\'{u}}ndez, Pernot, Cuppen, Loison, \&
  Talbi}]{10WaSmHe}
Wakelam, V., Smith, I. W.~M., Herbst, E., {et~al.} 2010, Space Sci. Rev., 156,
  13

\bibitem[{{Wakelam} {et~al.}(2017){Wakelam}, {Bron}, {Cazaux}, {Dulieu}, {Gry},
  {Guillard}, {Habart}, {Hornek{\ae}r}, {Morisset}, {Nyman}, {Pirronello},
  {Price}, {Valdivia}, {Vidali}, \& {Watanabe}}]{17WaBrCa}
{Wakelam}, V., {Bron}, E., {Cazaux}, S., {et~al.} 2017, Mol. Astrophys, 9, 1

\bibitem[{{Wang} {et~al.}(2020){Wang}, {Li}, {Goldsmith}, {Zhang}, {Gao},
  {Shi}, {Li}, {Fang}, {Li}, \& {Zhang}}]{20WaLiGo}
{Wang}, J., {Li}, D., {Goldsmith}, P.~F., {et~al.} 2020, \apj, 889, 129

\bibitem[{{Wang} {et~al.}(2016){Wang}, {Tian}, {Li}, \& {Hu}}]{16WaTiLi}
{Wang}, Y., {Tian}, F., {Li}, T., \& {Hu}, Y. 2016, \icarus, 266, 15

\bibitem[{Webster {et~al.}(2015)Webster, Mahaffy, Atreya, Flesch, Mischna,
  Meslin, Farley, Conrad, Christensen, Pavlov, Mart{\'{i}}n-Torres, Zorzano,
  McConnochie, Owen, Eigenbrode, Glavin, Steele, Malespin, Archer, Sutter,
  Coll, Freissinet, McKay, Moores, Schwenzer, Bridges, Navarro-Gonzalez,
  Gellert, \& Lemmon}]{14WeMaAt}
Webster, C.~R., Mahaffy, P.~R., Atreya, S.~K., {et~al.} 2015, Science, 347, 415

\bibitem[{Western(2017)}]{16WeXXXX}
Western, C.~M. 2017, \jqsrt, 186, 221

\bibitem[{Western {et~al.}(2018)Western, Carter-Blatchford, Crozet, Ross,
  Morville, \& Tokaryk}]{18WeCaCr}
Western, C.~M., Carter-Blatchford, L., Crozet, P., {et~al.} 2018, \jqsrt, 219,
  127

\bibitem[{Whitworth \& Jaffa(2018)}]{18WhJaXX}
Whitworth, A.~P., \& Jaffa, S.~E. 2018, A\&A, 611, A20

\bibitem[{Willetts {et~al.}(1990)Willetts, Handy, Green, \&
  Jayatilaka}]{90WiHaGr}
Willetts, A., Handy, N.~C., Green, W.~H., \& Jayatilaka, D. 1990, J. Chem.
  Phys., 94, 5608

\bibitem[{Wolniewicz {et~al.}(1998)Wolniewicz, Simbotin, \&
  Dalgarno}]{98WoSiDa}
Wolniewicz, L., Simbotin, I., \& Dalgarno, A. 1998, ApJS, 115, 293

\bibitem[{Wordsworth \& Pierrehumbert(2014)}]{14WoPiXX}
Wordsworth, R., \& Pierrehumbert, R. 2014, \apjl, 785, 2

\bibitem[{{Yong} {et~al.}(2003){Yong}, {Lambert}, \& {Ivans}}]{yong:03}
{Yong}, D., {Lambert}, D.~L., \& {Ivans}, I.~I. 2003, \apj, 599, 1357

\bibitem[{Yu {et~al.}(2014)Yu, Drouin, \& Miller}]{14YuDrMi}
Yu, S., Drouin, B.~J., \& Miller, C.~E. 2014, J. Chem. Phys, 141, 174302

\bibitem[{Yurchenko {et~al.}(2009)Yurchenko, Barber, Yachmenev, Thiel, Jensen,
  \& Tennyson}]{09YuBaYa}
Yurchenko, S.~N., Barber, R.~J., Yachmenev, A., {et~al.} 2009, J. Phys. Chem.
  A, 113, 11845

\bibitem[{Yurchenko \& Tennyson(2014)}]{14YuTeXX}
Yurchenko, S.~N., \& Tennyson, J. 2014, \mnras, 440, 1649

\bibitem[{{Zapata Trujillo} {et~al.}(2021){Zapata Trujillo}, Syme, Rowell,
  Burns, Clark, Gorman, Jacob, Kapodistrias, Kedziora, Lempriere, Medcraft,
  O'Sullivan, Robertson, Soares, Steller, Teece, Tremblay, Sousa-Silva, \&
  McKemmish}]{21TrSyRo}
{Zapata Trujillo}, J.~C., Syme, A.-M., Rowell, K.~N., {et~al.} 2021, Front.
  Astron. Space Sci., 8, 1

\bibitem[{{Zhang} {et~al.}(2018){Zhang}, {Romano}, {Ivison}, {Papadopoulos}, \&
  {Matteucci}}]{zhang:18}
{Zhang}, Z.-Y., {Romano}, D., {Ivison}, R.~J., {Papadopoulos}, P.~P., \&
  {Matteucci}, F. 2018, \nat, 558, 260

\end{thebibliography}



\appendix

\section{Further Derivations}
\label{sec:Extra}

\subsection{Deriving the Energy}
\label{appendix:energy_deriv}
We will go through how to derive the corresponding energy for each vibrational level as referenced by equations \ref{eq:3} and \ref{eq:5}. We will use the following notation from section \ref{subsection:rot_constants} within the following appendix sections. As a reminder, the relation between the anharmonic position, $x$, and the harmonic position, $u$, can be described as:
\begin{equation}
    x-\sigma = u
\end{equation}

We will now demonstrate deriving the ground state energy. To begin we must use perturbation theory. As the anharmonic Hamiltonian (equation \ref{eq:1}) and wavefunctions are known (equation \ref{eq:2}) we can express it as:

\begin{equation}
    E_0 =  \braket{\psi_0^{\ast} | \hat{H} | \psi_0}
\end{equation}    
\begin{equation}
    = \int_{-\infty}^{\infty} \psi_0^{\ast} \left(-\frac{1}{2}\frac{\partial^2}{\partial x^2} + \frac{1}{2!}\eta_{ii}x^2 + \frac{1}{3!}\eta_{iii}x^3 + \frac{1}{4!}\eta_{iiii}x^4\right)\psi_0\, dx
\end{equation}

For ease of comprehension we will break the following derivations down into four components and solve individual, then sum back together. We will now describe the separated parts and label for ease:
\begin{align}
    A &= \int_{-\infty}^{\infty} \psi_0^{\ast} \left(-\frac{1}{2}\frac{\delta^2}{\delta x^2}\right) \psi_0\, dx 
    \\
    B &= \int_{-\infty}^{\infty}\psi_0 \left( \frac{1}{2!}\eta_{ii}x^2\right)\psi_0\, dx
    \\
    C &= \int_{-\infty}^{\infty}\psi_0 \left( \frac{1}{3!}\eta_{iii}x^3\right)\psi_0\, dx
    \\
    D &= \int_{-\infty}^{\infty}\psi_0 \left( \frac{1}{4!}\eta_{iiii}x^4\right)\psi_0\, dx
\end{align}{}

\indent Lets begin with A first, and we need to remember that when an operator involves a differentiation, it does not commute. This means we now need to first differentiate $\psi_0$ twice, but leave $\psi_0$ unaffected:
\begin{equation}
    A = \int_{-\infty}^{\infty} \psi_0^{\ast} \left(-\frac{1}{2}\frac{\partial^2\psi_0}{\partial Q^2}\right)\, dx
\end{equation}
Now to insert the equations for the wavefunctions in:
\begin{align}
    &= -\frac{1}{2} \left(\frac{\omega}{\pi}\right)^{1/2} \int_{-\infty}^{\infty} e^{-\frac{\omega}{2}\left(x-\sigma\right)^2}\frac{\partial^2}{\partial x^2}e^{-\frac{\omega}{2}\left(x-\sigma\right)^2}\, dx
\end{align}
Recalling $u=x-\sigma$ therefore, $dx = du$. Substitute this back in:
\begin{align}
    &= -\frac{1}{2} \left(\frac{\omega}{\pi}\right)^{1/2} \int_{-\infty}^{\infty} e^{-\frac{\omega}{2}\left(u\right)^2}\frac{\partial^2}{\partial x^2}e^{-\frac{\omega}{2}\left(u\right)^2}\, du{}
\end{align}
Doubly differentiating within the equation we get:
\begin{align}
     \frac{\partial}{\partial x}e^{-\frac{\omega}{2}\left(u\right)^2}
    &= -\omega ue^{-\frac{\omega}{2}\left(u\right)^2} {}
    \\
    \frac{\partial}{\partial x} -\omega ue^{-\frac{\omega}{2}\left(u\right)^2} &= u^2\omega^2 e^{-\frac{\omega}{2}\left(u\right)^2} - \omega e^{-\frac{\omega}{2}\left(u\right)^2}
\end{align}
\begin{align}
    \therefore &= -\frac{1}{2} \left(\frac{\omega}{\pi}\right)^{1/2} \int_{-\infty}^{\infty} e^{-\frac{\omega}{2}\left(u\right)^2} \left(u^2\omega^2 e^{-\frac{\omega}{2}\left(u\right)^2} - \omega e^{-\frac{\omega}{2}\left(u\right)^2}\right)\, du{}
    \\
    &= -\frac{1}{2} \left(\frac{\omega}{\pi}\right)^{1/2} \int_{-\infty}^{\infty} \left(u^2\omega^2 e^{\omega\left(u\right)^2} - \omega e^{-\omega\left(u\right)^2}\right)\, du{}
    \\
    &= -\frac{1}{2} \left(\frac{\omega}{\pi}\right)^{1/2}  \left[\omega^2 \frac{1}{2\omega}\left(\frac{\pi}{\omega}\right)^{1/2} - 2\omega\left(\frac{\pi}{4\omega}\right)^{1/2}\right]{}
    \\
    &= \frac{1}{4}\omega
\end{align}

Now onto part B, same as before we will introduce the wavefunctions then use substitution to solve:
\begin{align}
    B &= \int_{-\infty}^{\infty}\psi_0^\ast \left( \frac{1}{2!}\eta_{ii}x^2\right)\psi_0\, dx
    \\
    &= \frac{1}{2!}\eta_{ii}\left(\frac{\omega}{\pi}\right)^{1/2}\int_{-\infty}^{\infty} x^2 e^{-\omega(x-\sigma)^2}\, dx
    \\
    &= \frac{1}{2!}\eta_{ii}\left(\frac{\omega}{\pi}\right)^{1/2}\int_{-\infty}^{\infty} (u + \sigma)^2 e^{-\omega u^2}\, du
    \\
    &= \frac{1}{2!}\eta_{ii}\left(\frac{\omega}{\pi}\right)^{1/2}\int_{-\infty}^{\infty} (u^2 + 2u\sigma + \sigma^2)e^{-\omega u^2}\, du
\end{align}
Once again we can break this down into pieces:
\begin{align}
    &\int_{-\infty}^{\infty} (u^2)e^{-\omega u^2}\, du = \frac{1}{2}\left(\frac{\pi}{\omega^3}\right){}
    \\
    &2\sigma \int_{-\infty}^{\infty} (u)e^{-\omega u^2}\, du = 0{}
    \\
    &\sigma^2\int_{-\infty}^{\infty} e^{-\omega u^2}\, du = \sigma^2\left(\frac{\pi}{\omega}\right)
\end{align}
Putting these components back together:
\begin{align}
    &= \frac{1}{2!}\eta_{ii} \left(\frac{\omega}{\pi}\right)^{1/2}\left(\frac{1}{2\omega}\left(\frac{\pi}{\omega}\right)^{1/2}+\sigma^2 \left(\frac{\pi}{\omega}\right)^{1/2}\right)
    \\
    &= \frac{1}{2!}\eta_{ii} \left[\frac{1}{2\omega} + \sigma^2\right]
\end{align}

Part C, using same method as before:
\begin{align}
    C &= \int_{-\infty}^{\infty}\psi_0^\ast \left( \frac{1}{3!}\eta_{iii}x^3\right)\psi_0\, dx
    \\
    &= \frac{1}{3!}\eta_{iii} \left(\frac{\omega}{\pi}\right)^{1/2} \int_{-\infty}^{\infty} x^3e^{-\omega(Q-\sigma)^2}\, dx
    \\
    &= \frac{1}{3!}\eta_{iii} \left(\frac{\omega}{\pi}\right)^{1/2} \int_{-\infty}^{\infty} (u+\sigma)^3e^{-\omega(u)^2}\, du
    \\
     &= \frac{1}{3!}\eta_{iii} \left(\frac{\omega}{\pi}\right)^{1/2} \int_{-\infty}^{\infty} (u^3 + 3\sigma u^2 + \sigma^2u + \sigma^3)^3e^{-\omega(u)^2}\, du
\end{align}
Breaking it down into parts again for ease of solving:
\begin{align}
    &\int_{-\infty}^{\infty} u^3 e^{-\omega(u)^2}\, du = 0
    \\
    3\sigma &\int_{-\infty}^{\infty} u^2 e^{-\omega(u)^2}\, du = 3\sigma \left(\frac{\pi}{4\omega^3}\right)^{1/2}
    \\
    \sigma^2 &\int_{-\infty}^{\infty} u e^{-\omega(u)^2}\, du = 0
    \\
    \sigma^3 &\int_{-\infty}^{\infty} e^{-\omega(u)^2}\, du = \sigma^3 \left(\frac{\pi}{\omega}\right)^{1/2}
\end{align}
Substituting these components back into the main equation:
\begin{align}
    &= \frac{1}{3!}\eta_{iii}\left(\frac{\omega}{\pi}\right)^{1/2}\left[\frac{3\sigma}{2\omega}\left(\frac{\pi}{\omega}\right)^{1/2} + \sigma^3\left(\frac{\pi}{\omega}\right)^{1/2}\right]
    \\
    &= \frac{1}{3!}\eta_{iii}\left[\frac{3\sigma}{2\omega} + \sigma^3\right]
\end{align}

Finally onto section D of the original equation:
\begin{align}
    D &= \int_{-\infty}^{\infty}\psi_0^\ast \left( \frac{1}{4!}\eta_{iiii}x^4\right)\psi_0\, dx
    \\
    &= \frac{1}{4!}\eta_{iiii} \left(\frac{\omega}{\pi}\right)^{1/2} \int_{-\infty}^{\infty} x^4 e^{-\omega(Q-\sigma)^2}\, dx
    \\
    &= \frac{1}{4!}\eta_{iiii} \left(\frac{\omega}{\pi}\right)^{1/2} \int_{-\infty}^{\infty} (u+\sigma)^4 e^{-\omega(u)^2}\, du
    \\
    &= \frac{1}{4!}\eta_{iiii} \left(\frac{\omega}{\pi}\right)^{1/2} \int_{-\infty}^{\infty} \left(u^4 + 4\sigma u^3 + 6\sigma^2 u^2 + 4\sigma^3 u + u^4\right)e^{-\omega u^2}\, du
\end{align}
Break down into parts:
\begin{align}
    &\int_{-\infty}^{\infty} u^4 e^{-\omega u^2} du = \frac{6}{8\omega^2}\left(\frac{\pi}{\omega}\right)^{1/2}
    \\
    &4\sigma \int_{-\infty}^{\infty} u^3 e^{-\omega u^2} du = 0
    \\
    &6\sigma^2\int_{-\infty}^{\infty} u^2 e^{-\omega u^2} du = \frac{6\sigma^2}{2\omega}\left(\frac{\pi}{\omega}\right)^{1/2}
    \\
    &4\sigma^3 \int_{-\infty}^{\infty} u\, du = 0
    \\
    & \sigma^4 \int_{-\infty}^{\infty} e^{-\omega u^2} du = \sigma^4 \left(\frac{\pi}{\omega}\right)^{1/2}
\end{align}
Summing the piece back together:
\begin{align}
    &= \frac{1}{4!}\eta_{iiii} \left(\frac{\omega}{\pi}\right)^{1/2} \left[\frac{6}{8\omega^2}\left(\frac{\pi}{\omega}\right)^{1/2} + \frac{6\sigma^2}{2\omega}\left(\frac{\pi}{\omega}\right)^{1/2} + \sigma^4 \left(\frac{\pi}{\omega}\right)^{1/2}\right]
    \\
    &= \frac{1}{4!}\eta_{iiii} \left[\frac{3}{4\omega^2} + \frac{3\sigma^2}{\omega} + \sigma^4\right]
\end{align}{}

\indent If we now take all the parts, A, B, C and D we can now have the full equation for the energy of the ground state, which matches the expression in equation \ref{eq:6}:
\begin{equation}
    E_0 = \frac{1}{4}\omega
    +\frac{1}{2!}\eta_{ii} \left[\frac{1}{2\omega} + \sigma^2\right]
    +\frac{1}{3!}\eta_{iii}\left[\frac{3\sigma}{2\omega} + \sigma^3\right]
    +\frac{1}{4!}\eta_{iiii} \left[\frac{3}{4\omega^2} + \frac{3\sigma^2}{\omega} + \sigma^4\right]
\end{equation}{}
The method described here can be applied to the first excited state to achieve the answer shown in equation \ref{eq:6}.

\subsection{Position Expectation Values}
\label{appendix:posexpval}
Let's first start with expectation value for the harmonic position, $u$:
\begin{equation}
    \braket{u} = \int_{-\infty}^{\infty} u |\psi|^2 du
\end{equation}
We can use a single simple identity to calculate the expectation value, as we know the wavefunction can be normalised:
\begin{equation}
    \int_{-t}^{t} o(u) du = 0 \,;\, |\psi|^2 = 1
\end{equation}
\begin{equation}
    \therefore \braket{u} = 0
\end{equation}

Now to do the anharmonic position expectation values. Although the anharmonic wavefunctions are also normalised we cannot use only the identities described previously, as the equations are more complex. Therefore, in this case we need to define the wavefunctions. Using \ref{eq:2}, we see that the for the ground state is:
\begin{equation}
    \psi_0 = \left(\frac{\omega}{\pi}\right)^{1/4} e^{-\frac{\omega}{2}(x - \sigma)^2}
\end{equation}

Additionally for the first excited vibrational state:
\begin{equation}
    \psi_1 = \left(\frac{4\omega^3}{\pi}\right)^{1/4}
    (x-\sigma) e^{-\frac{\omega}{2}(x - \sigma)^2}
\end{equation}

Now to derive the expectation value for the shifted anharmonic position, $\braket{x}$, with respect to the anharmonic wavefunctions. We can now derive the expectation value of the position, $x$, for the anharmonic ground and first vibrational level. 
\begin{equation}
    \braket{\psi_0|x|\psi_0} = \int_{-\infty}^{\infty} |\psi_0|^2 x\, dx
\end{equation}
\begin{equation}
    = \left(\frac{\omega}{\pi}\right)^{1/2} \int_{-\infty}^{\infty} x e^{-\omega(x - \sigma)^2}\, dx
\end{equation}
Substituting $x = u + \sigma$:
\begin{equation}
    = \left(\frac{\omega}{\pi}\right)^{1/2} \int_{-\infty}^{\infty} (u+\sigma) e^{-\omega(u)^2}\, du
\end{equation}
Separating into two components (A and B) for easier integration:
\begin{equation}
    A = \int_{-\infty}^{\infty} (u)e^{-\omega(u)^2}\, du
\end{equation}
Using the identity for an odd function (where $o(u)=-o(-u)$ for all u), and knowing the wavefunction can be normalised:
\begin{equation}
    \int_{-t}^{t} o(u) du = 0
\end{equation}
\begin{equation}
    \therefore A = \int_{-\infty}^{\infty} (u)e^{-\omega(u)^2}\, du = 0
\end{equation}
Now for part B:
\begin{equation}
    B = \sigma \int_{-\infty}^{\infty} e^{-\omega(u)^2}\, du
\end{equation}
An even function $e(u)$ satisfies $e(u) = e(-u)$ for all $u$. There, for any $t$,
\begin{equation}
    \int_{-t}^{t} e(u) du = 2\int_{0}^{t} e(u) du
\end{equation}
\begin{equation}
\therefore B = \sigma \int_{-\infty}^{\infty} e^{-\omega(u)^2}\, du = 2\sigma \int_{0}^{\infty} e^{-\omega(u)^2}\, du 
\end{equation}
Then using a secondary integration identity:
\begin{equation}
   \int_{0}^{\infty} e^{- \omega u^2} du = \left(\frac{\pi}{4\omega}\right)^{1/2}
\end{equation}
\begin{equation}
    \therefore B = 2\sigma \left(\frac{\pi}{4\omega}\right)^{1/2}
\end{equation}
Put A and B back together into the original expression:
\begin{equation}
     = \left(\frac{\omega}{\pi}\right)^{1/2} \left[0 + 2\sigma\left(\frac{\pi}{4\omega}\right)^{1/2}\right]
\end{equation}
\begin{equation}
    = \sigma 
\end{equation}

Now onto the first vibrational level:
\begin{equation}
    \braket{\psi_1|x|\psi_1} = \int_{-\infty}^{\infty} |\psi_1|^2 x\, dx
\end{equation}
\begin{equation}
    = \left(\frac{4\omega^3}{\pi}\right)^{1/2} \int_{-\infty}^{\infty} (x - \sigma)^2 x e^{-\omega(x - \sigma)^2}\, dx
\end{equation}
Substituting $x = u + \sigma$:
\begin{equation}
    = \left(\frac{4\omega^3}{\pi}\right)^{1/2} \int_{-\infty}^{\infty} u^2(u+\sigma) e^{-\omega(u)^2}\, du
\end{equation}
\begin{equation}
    = \left(\frac{4\omega^3}{\pi}\right)^{1/2} \int_{-\infty}^{\infty} (u^3+\sigma u^2) e^{-\omega(u)^2}\, du
\end{equation}
Once again separating into two components (A and B) for easier integration:
\begin{equation}
    A = \int_{-\infty}^{\infty} (u^3)e^{-\omega(u)^2}\, du
\end{equation}
Using the identity for an odd function (where $o(u)=-o(-u)$ for all u):
\begin{equation}
    \int_{-t}^{t} o(u) du = 0
\end{equation}
\begin{equation}
    \therefore A = \int_{-\infty}^{\infty} (u^3)e^{-\omega(u)^2}\, du = 0
\end{equation}
\begin{equation}
    B = \sigma \int_{-\infty}^{\infty} (u^2)e^{-\alpha(u)^2}\, du 
\end{equation}
Using the integration identity:
\begin{equation}
   \int_{-\infty}^{\infty} u^2e^{-\omega u^2} du = \frac{1}{2} \left(\frac{\pi}{\omega^3}\right)^{1/2}
\end{equation}
\begin{equation}
    \therefore B = \frac{\sigma}{2}\left(\frac{\pi}{\omega^3}\right)^{1/2}
\end{equation}
Put A and B back together into the original expression:
\begin{equation}
    = \left(\frac{4\omega^3}{\pi}\right)^{1/2} \left[0 + \frac{\sigma}{2}\left(\frac{\pi}{\omega^3}\right)^{1/2}\right]
\end{equation}
\begin{equation}
    = \sigma
\end{equation}

Clearly the outcome is independent of the vibrational level.

\subsection{Branches}
\label{appendix:branches}
This next section will detail the working out required to produce equations \ref{eq:26}, \ref{eq:27}, \ref{eq:28}, \ref{eq:29} and \ref{eq:30}.
The general analytical expression for a spectrum of a diatomic molecule is the following:
\begin{align}
    \varepsilon_{(J, v)} &= \varepsilon_J + \varepsilon_v
    \\
    &= \bar{\omega}_v(v+1/2) + B_vJ(J+1) - D_eJ^2(J+1)^2
\end{align}
Where:
\begin{align}
    \bar{\omega}_v &= \bar{\omega}_e - \bar{\omega}_ex_e(v+1/2){}
    \\
    \therefore \varepsilon_{(J, v)} &= \bar{\omega}_e(v+1/2) - \bar{\omega}_ex_e(v+1/2)^2 + B_vJ(J+1) - D_eJ^2(J+1)^2 + ...
\end{align}

The frequency $\nu_{0\rightarrow 1}$ is usually called the band origin or band centre. Which is represented by the following equation:
\begin{equation}
    \nu_{0\rightarrow 1} = \bar{\omega}_e - 2\bar{\omega}_ex_e 
\end{equation}
$D_e$ represents the centrifugal distortion constant.
For current theory we can ignore the small centrifugal distortions from $D_e$ etc, and the equation can be rewritten simply as:
\begin{equation}
    \varepsilon_{(J, v)} = \bar{\omega}_e(v+1/2) - \bar{\omega}_ex_e(v+1/2)^2 + B_vJ(J+1)
\end{equation}

For the next part we will be restricting the discussion to solely fundamental vibrational transitions, $v=0\rightarrow v=1$. We also take the respective B values as $B_0$ and $B_1$ with $B_0>B_1$. This transition can generally be described as (in $cm^{-1}$):
\begin{align}
    \Delta\varepsilon &= \varepsilon_{J',v=1} - \varepsilon_{J'',v=0}
    \\
    &= \nu_{0\rightarrow 1} + B_1J'(J'+1) - B_0J''(J''+1)
\end{align}

The transitions for the P Branch ($\Delta J = -1$, $J''=J'+1$) is given by:
\begin{align}
    \Delta\varepsilon &= \bar{\nu}_P 
    \\
    &= \nu_{0\rightarrow 1} - (B_1 + B_0)(J'+1) + (B_1 - B_0)(J'+1)^2
\end{align}
Where; $J'=0,1,2,3,..$.

The transitions for the Q Branch ($\Delta J = 0$, $J''=J'$) is given by:
\begin{align}
    \Delta\varepsilon &= \bar{\nu}_Q 
    \\
    &= \nu_{0\rightarrow 1} - J'(J'+1)(B_1 - B_0)
\end{align}
Where; $J'=0,1,2,3,..$.

Therefore for the R Branch ($\Delta J = +1$, $J'=J''+1$) the transition is given as:
\begin{align}
    \Delta\varepsilon &= \bar{\nu}_R 
    \\
    &= \nu_{0\rightarrow 1}
    + (B_1 + B_0)(J''+1) + (B_1 - B_0)(J''+1)^2
\end{align}
Where; $J''=0,1,2,3,...$ .

Taking the general rovibrational equation:
\begin{equation}
    \varepsilon(J, v) = \bar{\omega}_e(v+1/2) - \bar{\omega}x_e(v+1/2)^2 + B_vJ(J+1)
\end{equation}
Lets now apply this for the ground state with quantum number J'' and the upper vibrational state with quantum number J' as per standard notation.
\begin{align}
    \varepsilon_{(J'', v=0)} &= \frac{1}{2}\bar{\omega}_e - \frac{1}{4}\bar{\omega}x_e + B_0J''(J''+1){}
    \\
    \varepsilon_{(J', v=1)} &= \frac{3}{2}\bar{\omega}_e - \frac{9}{4}\bar{\omega}x_e + B_1J'(J'+1){}
    \\
    \therefore \Delta\varepsilon&= \nu_{0\rightarrow 1} + B_1J'(J'+1) - B_0J''(J''+1)
\end{align}

Now consider the S branch where $\Delta J = +2,\,J'=J''+2$. By plugging in these formulas:
\begin{align}
    \Delta\varepsilon &= \nu_S
    \\
    &= \nu_{0\rightarrow 1} + B_1(J''+2)(J''+3) - B_0J''(J''+2)
    \\
    &= \nu_{0\rightarrow 1} + B_1((J'')^2 + 5J'' + 6) - B_0((J'')^2 + J'')
\intertext{In the cases where $B_0 = B_1 = B$, the expression simplifies to:}
    \nu_S &= \nu_{0\rightarrow 1} + B(4J'' + 6)
\end{align}

Finally onto the O branch where $\Delta J = -2,\,J''=J'+2$. By plugging in these formulas:
\begin{align}
    \Delta\varepsilon &= \nu_O
    \\
    &= \nu_{0\rightarrow 1} + B_1J'(J'+1) - B_0(J'+2)(J'+3)
    \\
    &= \nu_{0\rightarrow 1} + B_1((J')^2+J') - B_0((J')^2+5J'+6)
\intertext{In the cases where $B_0 = B_1 = B$, the expression simplifies to:}
    \nu_O &= \nu_{0\rightarrow 1} - B(4J'' + 6)
\end{align}


\end{document}